\documentclass[journal=tosc]{iacrtrans}

\usepackage{amsthm} 

\theoremstyle{definition}          
\usepackage{algorithm}
\usepackage{algpseudocode}
\usepackage{adjustbox}
\usepackage{threeparttable}
\usepackage{leftindex}
\usepackage{booktabs}
\usepackage{enumitem}
\usepackage{listings}
\usepackage{subcaption}

\author{Chao Yin\inst{1,2} \and Zunchen Huang\inst{2} \and Chenglu Jin\inst{2} \and Marten van Dijk\inst{1,2} \and Fabio Massacci\inst{1,3}}
\institute{
  Vrije University, Amsterdam, the Netherlands
  \and
  Centrum Wiskunde \& Informatica, Amsterdam, the Netherlands 
  \and
  University of Trento, Italy
}

\title{Function Recovery Attacks in Gate-Hiding Garbled Circuits using SAT Solving}

\begin{document}

\maketitle

\keywords{Gate-hiding garbled circuits\and SAT attack\and Reverse engineering}


\begin{abstract}\label{sec:abstract}
Semi-Private Function Evaluation (SPFE) enables joint computation while protecting both input data and the function itself. A practical instantiation is gate-hiding garbled circuits, which conceal gate functionalities while revealing circuit topology. Existing security definitions intentionally exclude leakage through topology, leaving its concrete impact on function privacy largely unexplored.

We present a SAT-based function-recovery attack that reconstructs hidden gate operations from a circuit's public topology under two attacker knowledge models. Our approach combines topology-preserving simplification theorems with a decomposition of the recovery task into smaller SAT queries, thereby reducing the candidate gate-type assignment space and improving recovery performance.

We evaluate the attack on ISCAS benchmarks, representative secure computation circuits, and fault-tolerant sensor fusion circuits under a 24-hour recovery budget. Compared to a baseline attack, the optimized version substantially reduces recovery time and, in some cases, completes recovery within the evaluation budget where the baseline does not.

Our results show that revealing circuit topology can materially assist recovery of hidden gate functionality, identifying topology as a security-relevant leakage channel in gate-hiding garbled circuits.

\end{abstract}

\newif\ifrebuttal
\rebuttalfalse

\newcounter{reviewer}
\setcounter{reviewer}{0}
\newcounter{comment}[reviewer]
\newcounter{editor}
\setcounter{editor}{0}
\newcounter{aecomment}[editor]
\renewcommand{\thereviewer}{\Alph{reviewer}}
\renewcommand{\thecomment}{\thereviewer.\arabic{comment}}
\renewcommand{\theeditor}{AE}
\renewcommand{\theaecomment}{\theeditor.\arabic{aecomment}}
\def\reviewer{%
	\refstepcounter{reviewer}%
	\section*{\textsf{Comments from Reviewer }\thereviewer}}
\def\comment{%
	\refstepcounter{comment}%
\vspace*{\baselineskip}\noindent{{\large\textbf{Comment}~\thecomment:~}}\ignorespaces}
\def\editor{%
	\refstepcounter{editor}%
	\section*{\textsf{Comments from Editor}}}
\def\aecomment{%
	\refstepcounter{aecomment}%
	\vspace*{\baselineskip}\noindent{\textbf{Comment}~\theaecomment}\ignorespaces}

\def\reply{\par\vspace*{0.25\baselineskip}\noindent\relax{\textcolor{blue}{\textbf{Reply: }\ignorespaces}}}


\def\reviewnotemulti#1{\ifrebuttal\expandafter\marginpar{#1}\fi}
\def\reviewnotemultiminor#1{\ifrebuttalminor\expandafter\marginpar{#1}\fi}

\long\def\fixed#1{\ifrebuttal{\color{red}#1}\else{#1}\fi}



\def\Encode#1#2#3#4{\ensuremath{#4 \Leftrightarrow {\sf EncodeCircuit}_{#1}\left(#2,#3\right)}}
\def\EncodeTopo#1#2{\ensuremath{EncodeCircuit}_{#1}\left(#2\right)}
\def\Encodepair#1#2#3#4{\ensuremath{{\sf EncCircuit}_{#1}\left(#2,#3,#4\right)}}
\def\Encodepairndss#1#2#3#4#5#6{\ensuremath{{\sf EncCircuit}_{#1}\left(#2,#3,#4,#5,#6\right)}}
\def\Encodepairmodelb#1#2#3#4#5{\ensuremath{{\sf EncCircuit}_{#1}\left(#2,#3,#4,#5\right)}}
\def\Topology{\ensuremath{\tau}}
\def\Oracle#1{\ensuremath{{\sf Oracle}_f\left(#1\right)}}
\def\OraclemodelB#1{\ensuremath{{\sf Oracle}_{f,\textbf{y}}\left(#1\right)}}
\def\sgateset#1{\textbf{S}_{#1}}

\def\Inputset{\ensuremath{I}}
\def\Outputset{\ensuremath{O}}
\def\inputwire#1{\ensuremath{\Inputset_{#1}}}
\def\outputwire#1{\ensuremath{\Outputset_{#1}}}

\def\inputvectorItem#1{\textbf{x}^{(#1)}}
\def\outputvectorItem#1{\textbf{z}^{(#1)}}
\def\inputvector{\textbf{x}}
\def\outputvector{\textbf{z}}
\def\fixedinput{\textbf{y}}
\def\equivfixedinput{\textbf{y}'}

\def\Rset{\mathcal{R}}
\def\Lset{\mathcal{L}}
\def\Sset{\mathcal{S}}
\def\Zset{\mathcal{Z}}
\def\gate#1{\mathsf{#1}}
\def\gateType#1{\mathsf{#1}}

\def\gateset{\mathcal{N}}
\def\gateItem#1{g_{#1}}
\def\leftDrivenGate{\gateItem{i}^{a}}
\def\rightDrivenGate{\gateItem{i}^{b}}
\def\gateInpWire#1#2{w_{#1}^{in_{#2}}}
\def\gateOutpWire#1#2{w_{#1}^{out_{#2}}}

\def\sgateset{\mathcal{G}_s}
\def\sgateItem#1{s_{#1}}
\def\zgateset{\mathcal{G}_z}
\def\zgateItem#1{z_{#1}}

\def\numofGates{k}
\def\numofS{l}
\def\circInputSize{n}
\def\fixedInputsize{n'}
\def\circOutputSize{m}
\def\disSetSize{r}
\def\originalCircuit{C}
\def\equivalentCircuit{C'}
\def\discriminateSet{DI}

\def\gateAssignment{\mathcal{T}}
\def\equivGateAssignment{\mathcal{T'}}

\def\typeVar#1{t_{#1}}

\def\evalcircuit#1{\text{Eval}(#1)}

\def\LabelList{{LabelList}}
\def\VariableGateAssignment{T}
\def\EncodeOneCircuit{F_{\mathrm{one}}}
\def\EncodeTwoCircuit{F_{\mathrm{two}}}
\def\BlockGateAssignment{\mathsf{BlockCirc}}
\def\FindDisInput{\mathsf{DiffConstr}}
\def\FindDisInputTwoCirc{\mathsf{DiffConstr}}
\def\EncodeFindDisInput{F_{dis}}
\def\FinalEncoding{F_{\mathrm{final}}}
\def\EncodeCircTopo{\mathsf{EncodeCircuit}}
\def\EncodeTwoCircTopo{\mathsf{EncodeTwoCircuit}}
\def\Allowed{D}
\def\Sel{\text{Sel}}
\def\classifyGates{\mathsf{ClassifyGates}}
\def\globalBlockF{F_{GBlock}}
\def\roundBlockF{F_{RBlock}}

\newcommand{\aeComment}[1]{}

\section{Introduction} \label{sec:intro}

In contrast to standard secure multi-party computation, which assumes the function to be public and only protects the parties' inputs, \emph{Private Function Evaluation} (PFE) aims to protect both the input data and the function being evaluated \cite{mohassel2013hide}. Typical PFE protocols transform a (private) circuit of size $n$ into a public universal circuit of size 
$\Theta(n \log n)$ \cite{kolesnikov2008practical,zhao2024improving}. Although asymptotically optimal, this expansion introduces substantial overhead. Fully homomorphic encryption approaches alternative to universal circuits~\cite{li2024privacy,li2024non,zhu2025private} have improved expressiveness, but inherit the computational cost of homomorphic encryption, including large ciphertexts, expensive key-switching and rotations, and bootstrapping~\cite{Shen2025BootstrappingSurvey,yu2026faster}. 

These computational hurdles have motivated a relaxation of PFE to a \emph{semi-private function evaluation} (Semi-PFE). It relaxes function privacy to allow limited structural leakage \cite{paus2009practical,gunther2019poster,gunther2024fluent,lin2023skipping}. A prominent instantiation of Semi-PFE is \emph{Gate-Hiding} (GH) garbled circuits, which intentionally reveal the circuit topology while hiding the type of each gate \cite{kempka2016circumvent,rosulek2017improvements,rosulek2021three,wang2024reducing}. Eventually, a unified construction was proposed that supports all two-input Boolean functions with near-optimal per-gate costs ~\cite{rosulek2021three}.

These efficiency gains hold under a security model that treats topology leakage as acceptable. \fixed{\reviewnotemulti{\ref{ra:weaknesses}}In hardware gate camouflaging, El Massad et al.~\cite{el2015integrated}  have been able to reverse engineer the camouflaged gate identities in the ISCAS benchmarks by using a SAT solver. Still, these attacks assume strong attacker knowledge: camouflaged gates are restricted to a small set of candidate functions, and \emph{the vast majority of gates are known}.} 

In contrast, \emph{GH garbled circuits hide all gates}, each of which may realize any of the sixteen two-input Boolean functions~\cite{rosulek2021three}. \fixed{\reviewnotemulti{\ref{ae:5}}As noted by Paus \cite{paus2009practical}\footnote{
    Clearly, [\ldots] if the number of PPBs [Privately Programmable Block - here Hidden Gates] $n$ or number of possible functionalities of PPBs $|Fi|$ is small, evaluator Alice might guess the correct function with high probability or probe the system via exhaustive search which must be prohibited by other means.}}, this appears to pose a very strong practical hurdle once the number of hidden gates becomes large.

If one considers the case of $n$-bit point functions (functions evaluating to one at exactly one point and to zero everywhere else), exhaustive search seems to be the obvious and only attack. How can one reverse engineer, with oracle access, a gate-hiding circuit implementing a point function out of the possible $2^{2^n}$ Boolean functions without tabulating all $2^n$ inputs?

This would be true if we did not know the topology, but we do. Knowing the topology changes the problem: instead of identifying an arbitrary Boolean function, the attacker only needs to determine which gate-type assignment on the known topology is consistent with the oracle responses. SAT constraints can then combine this structural information with the queried input-output (I/O) pairs to rule out inconsistent candidates and recover the target functionality with far fewer oracle queries than exhaustive truth-table evaluation. This makes function recovery surprisingly efficient in practice: for example, a 16-bit point function can be recovered with only 85 oracle queries instead of testing all 65,536 inputs. Likewise, over the complete set of 256 distinct 8-bit point functions, our SAT-based recovery attack successfully recovered every instance, requiring a median of 131 oracle queries and a minimum observed query count of 29, without exhaustively querying all 256 inputs.

\subsection{Our Contributions}\label{sec:contributions}
Our main goal is to understand how leakage of circuit topology can be exploited for function recovery. To this end, we show that topology alone constrains the space of candidate gate-type assignments, and that these structural constraints can be combined with SAT-based modeling and optimized recovery procedures to reduce recovery time.


\begin{trivlist}\setlength{\itemindent}{1.5em}

\item[(i)] We show that circuit topology alone imposes strong constraints on possible gate-type assignments. Using only the circuit topology, we can restrict the set of admissible gate types at each gate while preserving the overall input-output behavior of the circuit. As a result, for 
$k$ such gates, an adversary needs to consider only 
$8^k$, $6^k$ or $3^k$
 possible gate-type assignments, instead of the 
$16^k$
 gate-type assignments allowed by gate-hiding schemes~\cite{rosulek2021three}.
\item[(ii)] We develop the first SAT-based function recovery attack tailored to gate-hiding garbled circuits. Our attack extends SAT-based gate decamouflaging to the GH setting, in which the semantics of \emph{every} gate are hidden cryptographically and each gate may realize any of the sixteen two-input Boolean functions.

\item[(iii)] Under a fixed 24-hour evaluation budget on commodity AWS hardware, our attack recovers gate-type assignments consistent with the target functionality for a diverse set of circuits with hundreds of gates and, in some cases, hundreds of input bits, using a limited number of oracle queries. Although these circuit sizes remain below those of many real-world deployments, our results provide empirical evidence that circuit topology leakage can be exploited algorithmically to assist function recovery in gate-hiding schemes.


\end{trivlist}

\textbf{Paper organization.} We begin by defining the system and threat models (\S\ref{sec:threat model}) and 
formulate the function recovery problem (\S\ref{sec: function recovery}). 
We then present a set of topology-preserving simplification theorems that 
exponentially reduce the search space (\S\ref{sec:simplificationtheorems}), 
followed by our SAT-based formulation of the function recovery task 
(\S\ref{sec:SATEncoding}). Next, we describe the baseline attack, obtained by generalizing the decamouflaging attack proposed by El Massad et al~\cite{el2015integrated} to our setting (\S\ref{sec:baseline}), 
and introduce our optimized incremental variant (\S\ref{sec:advancedAttack}). 
We further extend the attack to the setting where the adversary controls only a 
subset of the inputs (\S\ref{sec: extended attack}). We then describe the circuit benchmarks used in our evaluation 
(\S\ref{sec:benchmarks}). Finally, we report experimental results 
(\S\ref{sec:implementation}), analyze point-function 
behavior (\S\ref{sec:pointfunctions}), present related work 
(\S\ref{sec:related_work}), discuss the limitations of the work (\S\ref{sec:limitations}), and conclude the paper (\S\ref{sec:conclusion}).


\textbf{Artifact Availability Statement.} Netlists of circuits, SAT encodings, Python code and experimental data will be  available on Zenodo upon acceptance.

\section{System Model and Threat Model}\label{sec:threat model}

\subsection{GH Semi-PFE System Model}
We consider a two-party Semi-PFE setting with a function owner Alice and an evaluator Bob. Alice holds a Boolean function $f$, represented as a Boolean circuit, while Bob holds a private input $x$. The goal is to allow Bob to obtain $f(x)$ without learning the function itself beyond the I/O behavior revealed through his evaluations. Bob knows the circuit topology of $f$, including the number of gates, their wiring, and the ordering of each gate's input wires, but not the type of any gate; each may implement any of the sixteen two-input Boolean functions~\cite{rosulek2021three}.

The computation is realized using GH garbled circuits. Alice garbles the circuit, preserving its topology while hiding all gate types, and sends the resulting garbled circuit to Bob. Bob obtains the garbled input labels corresponding to his private input via oblivious transfer and evaluates the garbled circuit gate by gate. At the end of the evaluation, Bob learns the output of the computation and nothing beyond what can be inferred from the output and the circuit topology.

\subsection{Threat Model}\label{subsec:threatmodelAB}
We consider Bob as an honest-but-curious adversary. That is, Bob follows the GH Semi-PFE protocol as prescribed, but tries to recover the hidden function $f$ from the information available to him during protocol execution.

As a legitimate evaluator, Bob has black-box oracle access to the function through repeated protocol executions. That is, Bob may adaptively choose inputs and observe the corresponding outputs, while having no access to intermediate wire values. We do not impose protocol-level restrictions on the number of evaluations; however, we consider only attacks that use a number of oracle queries that is significantly smaller than the full input space size $2^n$. The attack succeeds whenever the adversary can reconstruct any gate-type assignment whose resulting circuit is functionally equivalent to the original circuit.

Bob's private inputs are not revealed to Alice as input privacy is guaranteed by the underlying cryptographic protocol.

\noindent\textbf{Threat Model A (Full Input Control).}
In Threat Model A, Bob supplies inputs $x$ to the evaluation and observes the corresponding outputs $f(x)$. Bob obtains one or more I/O pairs $(x, f(x))$ and attempts to recover a gate-type assignment consistent with the observed oracle behavior.

\noindent\textbf{Threat Model B (Partial Input Knowledge).}
In Threat Model B, the target function takes two inputs and is written as $f(x,y)$, where $x$ is known to Bob and $y$ is an unknown but fixed secret throughout the attack. Bob knows the input domain of $y$, but not its value. Bob may adaptively choose inputs $x$ and observe the corresponding outputs $f(x,y)$. From Bob's perspective, this setting is equivalent to reverse-engineering a circuit with a fixed but unknown input.

\section{The Function Recovery Problem}\label{sec: function recovery}

We denote by $\mathcal{L}$ the library of all sixteen possible two-input Boolean gate types.
State-of-the-art gate-hiding schemes can encode any gate type in $\mathcal{L}$. \fixed{\reviewnotemulti{\ref{rc:DAGdefinition}}We model the circuit as a directed acyclic graph (DAG) with an ordering on incoming edges to distinguish the two input wires of each gate.}

\begin{definition}
\label{def:circuit}
Let $C = \{\gateset \cup \Inputset \cup \Outputset, E, \gateAssignment\}$ be the target circuit, where
$\gateset = \{g_0, g_1, \dots, g_{k-1}\}$ is the set of gate nodes,
$\Inputset$ denotes the set of external input nodes, and
$\Outputset$ denotes the set of output nodes.
The edge set
$E \subseteq (\Inputset \cup \gateset) \times (\gateset \cup \Outputset)$
specifies the directed wiring between nodes, and the mapping
$\gateAssignment : \gateset \rightarrow \mathcal{L}$
assigns each gate node a two-input Boolean gate type.
Let $\circInputSize = |\Inputset|$ and $m = |\Outputset|$.
Under the assignment $\gateAssignment$, the circuit computes a Boolean function
$\evalcircuit{C} : \{0,1\}^{\circInputSize} \rightarrow \{0,1\}^m$.
For an input vector $\inputvector \in \{0,1\}^{\circInputSize}$, we write
$\evalcircuit{C}(\inputvector) = \outputvector$
to denote the corresponding output vector $\outputvector \in \{0,1\}^m$.
\end{definition}

We denote by $\Topology(C) = (\gateset \cup \Inputset \cup \Outputset, E)$ the circuit topology, that is, the structure of the circuit without gate-type assignments.
Conversely, equipping a topology $\Topology(C)$ with a gate-type assignment $\gateAssignment$ yields a fully instantiated circuit $C$.
\fixed{\reviewnotemulti{\ref{rc:oracledefinition}}We denote black-box access to the circuit $C$ via $\Oracle{\cdot}$.
On input $\inputvector$, the oracle returns the corresponding output
$\outputvector$  $=\Oracle{\inputvector}$.}

We now formalize function recovery under Threat Model A: the attacker has adaptive black-box oracle access to the target circuit.

\begin{problem}[Function Recovery in Model A]
\label{prob:function-recovery-modelA}
Let the target Boolean circuit be $C = (\Topology, \gateAssignment)$, where
$\Topology = (\gateset \cup \Inputset \cup \Outputset, E)$.
The attacker is given the circuit topology $\Topology$, does not know the gate-type assignment
$\gateAssignment$, but may interact with the oracle $\Oracle{\cdot}$ adaptively. 

\textbf{Formal Goal.}
After adaptively querying the oracle on inputs
$\inputvectorItem{i} \in \{0,1\}^{\circInputSize}$ and receiving outputs
$\outputvectorItem{i} = \Oracle{\inputvectorItem{i}}$, the attacker aims to recover a gate-type assignment $\equivGateAssignment : \gateset \rightarrow \Lset$ such that the instantiated circuit
$\equivalentCircuit = (\Topology, \equivGateAssignment)$
is functionally equivalent to $C$; that is, $\forall \inputvector \in \{0,1\}^{\circInputSize} :
\evalcircuit{\equivalentCircuit}(\inputvector)
=
\evalcircuit{C}(\inputvector)=\Oracle{\inputvector}$.
Any such $\equivGateAssignment$ is considered a successful recovery.

\textbf{Query Efficiency Criterion.}
The attacker seeks such a recovery using a number of oracle queries that is significantly smaller than the exhaustive bound $2^{\circInputSize}$. We measure complexity primarily by the number of oracle queries 
$|\mathcal{Q}|$ and the computational cost required to derive $\equivGateAssignment$.

\end{problem}

We now consider Threat Model B, which extends the problem by assuming that the circuit input of total length $\circInputSize + \fixedInputsize$ is split into two parts. The adversary controls the input $\inputvector\in\{0,1\}^\circInputSize$ while the remaining $\fixedInputsize$-bit input $\fixedinput$ is fixed but unknown to the adversary. The circuit output remains $\circOutputSize$ bits.


\begin{problem}[Function Recovery in Model B]
\label{prob:function-recovery-modelB}
Let the target Boolean circuit be given by $C = (\Topology, \gateAssignment)$, where
$\Topology = (\gateset \cup \Inputset \cup \Outputset, E)$.
The adversary is given the circuit topology $\Topology$ but does not know the gate-type assignment
$\gateAssignment$ nor the value of the fixed input $\fixedinput$.
The adversary may interact with the oracle $\OraclemodelB{\cdot}$ adaptively.

\textbf{Formal Goal.}
The attacker adaptively queries the oracle on inputs $\inputvector \in \{0,1\}^{\circInputSize}$, where the oracle returns
$\OraclemodelB{\inputvector} = \evalcircuit{\originalCircuit}(\inputvector, \fixedinput) $ for a fixed but unknown $\fixedinput \in \{0,1\}^{\fixedInputsize}$. The goal is to recover a gate-type assignment $\equivGateAssignment : \gateset \rightarrow \Lset$
together with a candidate value $\equivfixedinput \in \{0,1\}^{\fixedInputsize}$ such that the instantiated circuit
$\equivalentCircuit = (\Topology, \equivGateAssignment)$ is functionally equivalent to the target circuit under the fixed input; that is,
$\forall \inputvector \in \{0,1\}^{\circInputSize}, \evalcircuit{\equivalentCircuit}(\inputvector, \equivfixedinput) = \evalcircuit{C}(\inputvector, \fixedinput)= \OraclemodelB{\inputvector}.$
\end{problem}

Multiple pairs $(\gateAssignment, \fixedinput)$ may induce the same functionality over the attacker-controlled input domain. Any such pair is considered a valid recovery.

\textit{Treatment of NOT Gates.}
\fixed{\reviewnotemulti{\ref{rb:notlocation},\ref{ae:notpoint}}Without loss of generality, we assume that NOT gates do not appear in the circuit.
Since $\Lset$ contains all sixteen two-input Boolean functions, any NOT gate can be absorbed into an adjacent two-input gate by complementing the corresponding wire value (including at primary inputs or outputs), without introducing new gates or changing the topology.
Accordingly, we focus on hiding and recovering the types of two-input gates.}

\section{Gate Search Space Simplification}\label{sec:simplificationtheorems}

Recovering an unknown Boolean circuit from oracle access can be viewed as a search problem over gate-type assignments that are consistent with the observed I/O behavior and the given topology.
For a circuit with $k$ gates drawn from the full two-input Boolean gate library $\Lset$ of $16$ primitive functions, the naive search space contains $|\Lset|^k = 16^k$ possible assignments that render exhaustive exploration infeasible, even for circuits of moderate size.


By analyzing local structural patterns of the revealed topology, and applying topology-preserving logical equivalences, we identify classes of gates whose admissible types can be restricted to much smaller subsets of $\Lset$ (e.g.\ of size $3$ or $6$) without affecting the circuit's overall I/O behavior. These restrictions lead to an exponential reduction in the adversary's search space.
Together, these topology-preserving simplifications provide a preprocessing step that reduces the candidate search space for SAT-based function recovery.

\subsection{Forward Propagation Reduction}\label{sec:R-wave}

\fixed{\reviewnotemulti{\ref{rb:algorithm:explanation},\ref{ra:weaknesses},\ref{ra:question},\ref{rc:15}}
The forward-propagation simplification exploits the fact that a Boolean negation at a gate's output can be pushed to its downstream consumers without changing the circuit's behavior or topology. An inverted output can be absorbed by reinterpreting the gate as its complemented variant (e.g., AND as NAND) and propagating the inversion forward. Iteratively applying this process pushes all internal negations to the primary outputs. Consequently, the type-selection domain is reduced from 16 to 8 for every gate whose output is not a primary output. Only gates whose outputs are primary outputs retain the full 16-type domain, since they cannot push negations to any subsequent gate.
}


\medskip




\begin{theorem}[R topology-preserving simplification]
\label{thm:R-wave}
Given any Boolean circuit
$\originalCircuit = (\gateset \cup \Inputset \cup \Outputset, E, \gateAssignment)$
over the full two-input gate-type library $\Lset$, there exists a gate-type
assignment $\equivGateAssignment$ such that the circuit
$\equivalentCircuit = (\gateset \cup \Inputset \cup \Outputset, E, \equivGateAssignment)$
is functionally equivalent to $\originalCircuit$, and every gate $g$ whose
output is not a primary output satisfies $\equivGateAssignment(g) \in \Rset$,
where
$$\Rset =
\{
\gate{XOR},\,
\gate{OR},\,
\gate{NAND},\,
\gate{TRUE},\,
\gate{\neg A},\,
\gate{\neg B},\,
\gate{(\neg A) \lor B},\,
\gate{A \lor (\neg B)}
\}.$$
For gates whose outputs are primary outputs, $\equivGateAssignment(g)$ may take
any value in $\Lset$. Moreover, for all $\inputvector \in \{0,1\}^{n}$,
$\evalcircuit{\originalCircuit}(\inputvector)
=
\evalcircuit{\equivalentCircuit}(\inputvector)$.
\end{theorem}



\subsection{Backward Propagation Reduction}\label{sec:szsimplification}

The second simplification exploits a complementary structural property of the circuit: some gates are the only successors of one or both of their predecessor gates. Such ``single-use'' wiring enables local functional rewrites that preserve topology and input--output behavior, while restricting the admissible gate types of the affected gates.




For example, if an $\gate{OR}$ gate is driven by two predecessor gates that both have fan-out = 1, then it can be rewritten as a
$\gate{NAND}$ gate by negating both of its inputs: $\gate{OR}(a,b) \;=\; \gate{NAND}(\neg a,\neg b).$ Since each predecessor gate has fan-out 1 and drives only this OR gate, these input negations can be absorbed into the predecessor gates without affecting any other part of the circuit.


\begin{definition}[S-Class Gate]\label{def:sclass}
A gate $g$ is an S-Class gate if:
(i) $g$ is not a primary-input gate, and 
(ii) both of its input-driving predecessor gates have fan-out equal to~$1$.
\end{definition}

\begin{definition}[Z-Class Gate]\label{def:zclass}
A gate $g$ is a Z-Class gate if:
(i) $g$ is not a primary-input gate, and 
(ii) exactly one of its input-driving predecessor gates has fan-out equal to~$1$.
\end{definition}






\begin{theorem} The sets of S-Class and Z-Class gates are disjoint.
\end{theorem}
\begin{theorem}[ZS topology-preserving simplification]\label{thm:sz}
Let $C = (\gateset\cup\Inputset\cup\Outputset,E,\gateAssignment)$ be a Boolean circuit over the full two-input gate-type library~$\Lset$. Then there exists a functionally equivalent circuit $C' = (\gateset\cup\Inputset\cup\Outputset,E,\gateAssignment')$ with the same node set and edge set such that $\forall \,\inputvector \in \{0,1\}^n, \,\evalcircuit{\originalCircuit}(\inputvector) = \evalcircuit{\equivalentCircuit}(\inputvector), $
and the following hold:
\begin{itemize}
    \item For every S-Class gate $g$, the assigned type satisfies $\gateAssignment'(g) \in \Sset = \{\gate{AND},\,\gate{NAND},\,\gate{XOR}\}$.
    
    \item For every Z-Class gate $g$, we refer to its two input wires as ``left'' and ``right'' merely following the netlist order, 
i.e., the first and second input wires in the Verilog description. If the unique fan-out\,{=}\,1 predecessor drives the left (resp.\ right) input of $g$, then
        $\gateAssignment'(g) \in \Zset_{\mathrm{left}} \ (\text{resp.} \ \Zset_{\mathrm{right}}),$ where
        $\Zset_{\mathrm{left}} = \{\gate{XOR},\,\gate{AND},\,\gate{NAND},\,\gate{NOR},\,\gate{OR},\,A\},\\[3pt]
          \Zset_{\mathrm{right}} = \{\gate{XOR},\,\gate{AND},\,\gate{NAND},\,\gate{NOR},\,\gate{OR},\,B\}.$

    \item All remaining gates (those not in S-Class or Z-Class) retain the full library: $\gateAssignment'(g) \in \Lset$.
    
\end{itemize}
\end{theorem}


The ZS simplification preserves both the circuit topology and its end-to-end Boolean behavior. 

\subsection{Combined Reduction (ZSR)}\label{sec:RSZ}



The simplifications in \S\ref{sec:R-wave} and~\S\ref{sec:szsimplification} exploit different circuit properties. The R simplification applies to all non-output-layer gates, while the ZS simplification applies only to Z-Class and S-Class gates. Since they target disjoint aspects of the circuit, the two are compatible and can be applied in either order as preprocessing before SAT solving.

In practice, however, combining the two simplifications requires limited bookkeeping to preserve their intended effects.
For example, when the ZS simplification is applied first, predecessor gates with fan-out $=1$ must be explicitly recorded and forced to retain the full gate-type library. When the R simplification is applied subsequently, these gates must be left unchanged.
This coordination ensures that previously imposed gate-type restrictions are respected, so that the resulting circuit preserves the same I/O behavior.

We now describe a concrete preprocessing procedure that applies the ZS simplification followed by the R simplification.

\textbf{Preprocessing procedure.}
Given a circuit $C = (\gateset\cup\Inputset\cup\Outputset,E,\gateAssignment)$, the attacker performs a single topological scan to obtain the fan-out of every gate and to identify the primary input and output layers. Each gate is classified as follows, and its admissible type set is then assigned accordingly:
\begin{enumerate}
    \item If both input-driving predecessors of a gate have fan-out $=1$, classify it as an S-Class gate and assign type set $\Sset$.

    \item Else if exactly one input-driving predecessor has fan-out $=1$, classify it as a Z-Class gate and assign type set $\Zset_{\mathrm{left}}$ or $\Zset_{\mathrm{right}}$ depending on whether the fan-out-1 predecessor drives the left or right input.
    
    \item Else if the gate is neither in the primary output layer nor an input-driving predecessor of any S- or Z-Class gate, assign type set $\mathcal{R}$.
    
    \item All remaining gates are classified as unrestricted and retain the full type set $\mathcal{L}$, as they may be required to absorb negations propagated from S/Z-Class gates or to handle residual negations at the output layer.
\end{enumerate}

\begin{table}[t]
\centering
\begin{tabular}{@{} l @{\hspace{1.5em}} l @{}}
\toprule
\textbf{Gate category} & \textbf{Admissible type set} \\
\midrule
S-Class gates & $\Sset$ (size $=3$) \\
Z-Class gates & $\Zset_{\mathrm{left}}$ or $\Zset_{\mathrm{right}}$ (size $=6$) \\
Gates assigned type set $\Rset$ & $\Rset$ (size $=8$) \\
Remaining gates & $\Lset$ (size $=16$) \\
\bottomrule
\end{tabular}
\caption{Resulting gate-type domains after applying the combined ZSR reduction.}
\label{tab:domains}
\end{table}

\begin{figure}[t]
  \centering
  \includegraphics[width=0.6\linewidth]{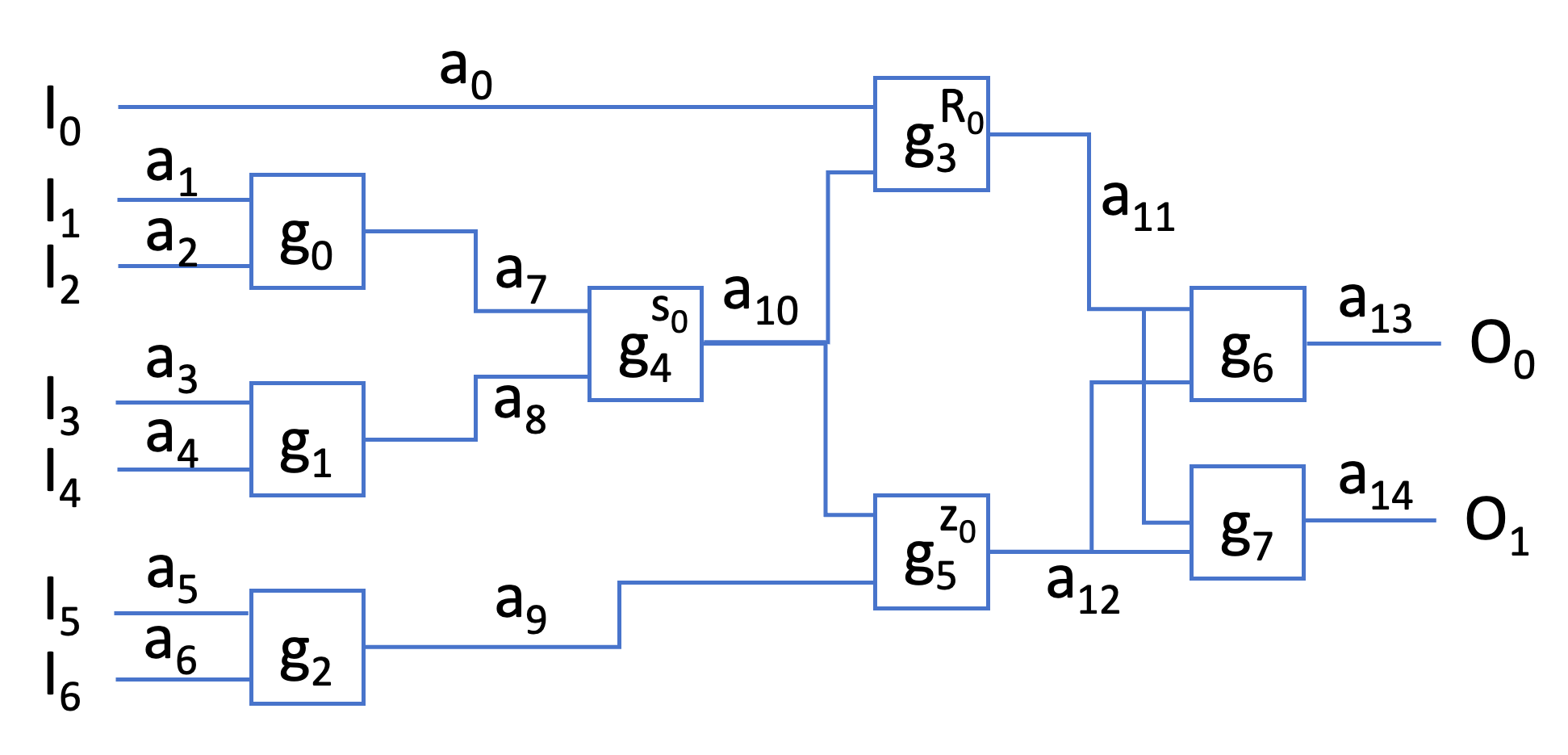}
  \caption{Example circuit annotated with S-Class gates ($S_0$), a Z-Class gate ($Z_0$), and gates assigned type set $\Rset$ ($R_0$). All remaining gates retain the full type library $\Lset$.}
  \label{fig:simplification-example}
\end{figure}

\textbf{Resulting restricted search space.}
\fixed{\reviewnotemulti{\ref{rd:5}}After this preprocessing step, the original gate-type assignment space
$|\Lset|^{|\gateset|}$ is reduced to $3^{|\gateset_S|} \times
    6^{|\gateset_Z|} \times
    8^{|\gateset_R|} \times
    16^{|\gateset_F|}$, where $\gateset_S$, $\gateset_Z$, and $\gateset_R$ denote the sets of S-Class gates, Z-Class gates, and gates assigned to type set $\Rset$, respectively.
The set $\gateset_F$ contains the remaining gates whose type domains are not reduced and therefore remain $\Lset$, including (i) primary output gates and (ii) non--S/Z-Class gates that directly precede an S- or Z-Class gate with fan-out 1.
For Z-Class gates, the admissible domain is $\Zset_{\mathrm{left}}$ or $\Zset_{\mathrm{right}}$, depending on the gate position.
Table~\ref{tab:domains} summarizes the resulting gate categories and their admissible type sets. }

\textbf{Example.}
Figure~\ref{fig:simplification-example} illustrates the combined gate-type domain reduction on a small circuit.
Gate $g_4 \in \gateset_S$ is classified as an S-Class gate because both of its input-driving predecessors $g_0$ and $g_1$ have fan-out $=1$, and its admissible domain is therefore restricted to $\Sset$.
Gate $g_5 \in \gateset_Z$ is classified as a Z-Class gate since exactly one of its input-driving predecessors ($g_2$) has fan-out $=1$; assuming wire $a_9$ feeds its left input, its type domain is restricted to $\Zset_{\mathrm{left}}$.
Gate $g_3 \in \gateset_R$ is assigned type set $\Rset$. 
All remaining gates ($g_0$, $g_1$, $g_2$, $g_6$, and $g_7$) belong to $\gateset_F$ and retain the full gate-type library $\Lset$.

In this example, the naive search space of $16^8\approx 4.3\cdot 10^9$ gate-type assignments is reduced to
$3^{1} \cdot 6^1 \cdot 8^1 \cdot 16^{5} \approx 1.5\cdot 10^8$, an order of magnitude reduction induced by topology-based restrictions.



\begin{remark}
\fixed{\reviewnotemulti{\ref{rd:nandnor},\ref{rd:9},\ref{ae:theoretical:scalability}}One may ask whether the restricted gate-type sets identified above can be further reduced to a basis consisting only of $\gate{NAND}$ or $\gate{NOR}$ gates.
Such a reduction is not possible \emph{without changing the topology}. The functional completeness of $\gate{NAND}$ and $\gate{NOR}$ relies on the ability to introduce additional gates and modify the circuit structure.
In contrast, our recovery model requires the circuit topology to be preserved: each gate must remain a single node with its original fan-in, fan-out, and position.
Under this constraint, each gate must be assigned a single primitive gate type rather than being implemented by a subcircuit.}
\end{remark}

The formal proofs of Theorems~\ref{thm:R-wave} and~\ref{thm:sz} are given in Appendix~\ref{Appendix: proofs}.

\section{SAT Encoding}\label{sec:SATEncoding}
\fixed{\reviewnotemulti{\ref{rb:algorithm:explanation},\ref{rc:paperwriting},\ref{rc:17},\ref{ae:7}}
We now turn to the SAT formulation of the recovery problem. Given a public circuit topology $\Topology$ and a set of observed input--output pairs $DI=\{(\inputvector,\outputvector)\}$, we ask whether there exists a gate-type assignment consistent with all observations and, if so, construct one. We encode this task using several CNF components, each capturing a different part of the recovery problem. Together, their satisfying assignments correspond to gate-type assignments on the known topology that reproduce the observed behavior.
}
Let $\Lset=\{f_0,\dots,f_{15}\}$ denote the full library of two-input Boolean functions. For each gate $g$ with unknown type, let $\Allowed(g)\subseteq\Lset$ denote its admissible type set. In the baseline formulation, $\Allowed(g)=\Lset$ for all gates. As discussed in Section~\ref{sec:simplificationtheorems}, preprocessing can restrict these admissible sets using topology alone, thereby reducing the SAT search space. We next define the Boolean variables, core constraints, and CNF encoding primitives used in our SAT-based function recovery attack.

Readers primarily interested in the high-level attack description may skip this section and proceed directly to Section~\ref{sec:baseline}.





\subsection{Variables and Basic Constraints}\label{sec:basicconstraints}

We begin by introducing the Boolean variables used in the SAT encoding and the basic constraints that relate them to the observed I/O behavior. Figure~\ref{fig:SATbasic} summarizes the variable set and constraint formulas.

\begin{figure}[t]
    \centering
    \small
    \setlength{\abovedisplayskip}{4pt}
    \setlength{\belowdisplayskip}{4pt}
    \setlength{\abovedisplayshortskip}{2pt}
    \setlength{\belowdisplayshortskip}{2pt}
    \caption{Boolean variables and basic constraints used in the SAT encoding.}
    \label{fig:SATbasic}
    \begin{align}
        \VariableGateAssignment
        \;:=\;&
        \{\Sel_{g,t}: g\in\gateset,\; t\in\Allowed(g)\} \label{equ:vgateassign}\\
        \Phi_{\mathrm{onehot}}(\VariableGateAssignment)
        \;:=\;&
        \bigwedge_{g\in\gateset}
        \left(
        \bigvee_{t\in\Allowed(g)} \Sel_{g,t}
        \;\wedge\;
        \bigwedge_{\substack{t,t'\in\Allowed(g),\, t\neq t'}}
        (\neg\Sel_{g,t}\vee\neg\Sel_{g,t'})
        \right) \label{equ:onehot}\\
        \Phi_{\text{s}}(\inputvector^{(i)},\outputvector^{(i)})
        \;:=\;&
        \bigwedge_{j=1}^{\circInputSize}\bigl(v^{(i)}_{\inputwire j}=\inputvector^{(i)}_j\bigr)
        \;\wedge\;
        \bigwedge_{h=1}^{\circOutputSize}\bigl(v^{(i)}_{\mathrm{pred}(\outputwire h)}=\outputvector^{(i)}_h\bigr)\label{equ:phis}\\
        \Phi_{\text{signals}}(\Topology,\discriminateSet)
        \;:=\;&
        \bigwedge_{(\inputvector^{(i)},\outputvector^{(i)})\in DI}
        \Phi_{\text{s}}(\inputvector^{(i)},\outputvector^{(i)})\label{equ:phisignals}
    \end{align}
\end{figure}

\noindent\textbf{Type-selection variables.}\quad
To encode an unknown gate assignment in SAT, we introduce the selector-variable family $\VariableGateAssignment$ defined in Equation~\eqref{equ:vgateassign}. Each variable $\Sel_{g,t}$ is true iff gate $g$ is assigned type $t$. Thus, $\VariableGateAssignment$ is the SAT-level representation of a candidate gate-type assignment, in contrast to the concrete mapping $\gateAssignment:\gateset\to\Lset$ in Definition~\ref{def:circuit}. Equation~\eqref{equ:onehot} enforces that exactly one admissible type is selected for each gate.

\noindent\textbf{Signal variables and boundary constraints.}\quad
For every observed sample $(\inputvector^{(i)},\outputvector^{(i)})\in DI$, we introduce Boolean signal variables for the node values induced by $\inputvector^{(i)}$. For every gate $g\in\gateset$, the variable $v^{(i)}_g$ denotes the Boolean value on the output of $g$, and for every primary input node $\inputwire j\in\Inputset$, the variable $v^{(i)}_{\inputwire j}$ denotes the corresponding input signal. As in Definition~\ref{def:circuit}, $\Inputset$ and $\Outputset$ denote the sets of primary input and output nodes, respectively, and each output node $\outputwire h\in\Outputset$ has a unique predecessor gate $\mathrm{pred}(\outputwire h)$ in the fixed topology $\Topology$. Equation~\eqref{equ:phis} enforces the boundary conditions for one sample by fixing the primary-input variables to $\inputvector^{(i)}$ and matching the output-driving gate signals to $\outputvector^{(i)}$; Equation~\eqref{equ:phisignals} lifts this constraint to all observed samples, yielding $\Phi_{\text{signals}}(\Topology,\discriminateSet)$.




\subsection{Circuit Encoding Primitives}
\label{subsec:circuitencoding}

We next define the CNF encoding primitives used in this work. Built from the variables and basic constraints introduced above, these encodings enforce consistency with the observed input--output pairs $\discriminateSet$, support the search for discriminating inputs, and exclude concrete assignments during the iterative recovery procedure. Figure~\ref{fig:SATencoding} summarizes the resulting primitives.



\begin{figure}[t]
    \centering
    \small
    \setlength{\abovedisplayskip}{4pt}
    \setlength{\belowdisplayskip}{4pt}
    \setlength{\abovedisplayshortskip}{2pt}
    \setlength{\belowdisplayshortskip}{2pt}
    \caption{SAT encoding primitives.}
    \label{fig:SATencoding}
    {\setlength{\jot}{2pt}
    \begin{align}
        \Phi_{\text{sem}}&(\Topology,\VariableGateAssignment,i)
        \;:=\;
        \bigwedge{}_{g\in\gateset}\bigwedge{}_{t\in\Allowed(g)}
        \Bigl(
        \Sel_{g,t}\Rightarrow\bigl(v^{(i)}_g\leftrightarrow f_t(v^{(i)}_u,v^{(i)}_w)\bigr)
        \Bigr) \label{equ:sem}\\
        \Phi_{\text{semantics}}&(\Topology,\VariableGateAssignment,\discriminateSet)
        \;:=\;
        \bigwedge{}_{(\inputvector^{(i)},\outputvector^{(i)})\in DI}
        \Phi_{\text{sem}}(\Topology,\VariableGateAssignment,i) \label{equ:semantics}\\
        \EncodeCircTopo&(\Topology,\VariableGateAssignment,\discriminateSet)
        \;:=\;
        \Phi_{\mathrm{onehot}}(\VariableGateAssignment)
        \wedge
        \Phi_{\text{signals}}(\Topology,\discriminateSet)
        \wedge
        \Phi_{\text{semantics}}(\Topology,\VariableGateAssignment,\discriminateSet) \label{equ:singlecirc}\\
        \EncodeTwoCircTopo&(\Topology,\VariableGateAssignment_1,\VariableGateAssignment_2,\discriminateSet)
        \;:=\;
        \bigwedge{}_{j\in\{1,2\}}        \EncodeCircTopo(\Topology,\VariableGateAssignment_j,\discriminateSet) \label{equ:twocirc}\\        
        \FindDisInputTwoCirc&(\Topology,\VariableGateAssignment_1,\VariableGateAssignment_2,X)
\;:=\;
\bigwedge_{j\in\{1,2\}}
\Bigl(
\bigwedge_{r=1}^{\circInputSize}\bigl(v^{(X)}_{\inputwire r,j}=X_r\bigr)
\wedge
\Phi_{\text{sem}}(\Topology,\VariableGateAssignment_j,X)
\Bigr)
\wedge
\nonumber\\&\bigvee_{h=1}^{\circOutputSize}
\Bigl(
v^{(X)}_{\mathrm{pred}(\outputwire h),1}\oplus
v^{(X)}_{\mathrm{pred}(\outputwire h),2}
\Bigr)\label{equ:finddis}\\        
        \BlockGateAssignment&(\VariableGateAssignment,\gateAssignment)
        \;:=\;        \bigvee{}_{g\in\gateset}\;\neg\Sel_{g,\,\gateAssignment(g)}
        \label{equ:block}
    \end{align}
    }
\end{figure}

\noindent\textbf{Gate semantics.}\quad
Equation~\eqref{equ:sem} gives the gate-semantics constraints for a single observed sample under the given circuit topology. It enforces that, under each selected gate type, the corresponding signal variables satisfy the Boolean semantics of that gate. For every gate $g\in\gateset$ with predecessors $u=\mathrm{pred}_L(g)$ and $w=\mathrm{pred}_R(g)$, the constraint ties the signal variable $v_g^{(i)}$ to the Boolean function $f_t(v_u^{(i)},v_w^{(i)})$ whenever $\Sel_{g,t}$ is true. Here $\Allowed(g)\subseteq\Lset$ denotes the admissible type set of $g$. Equation~\eqref{equ:semantics} aggregates these constraints over all observed samples in~$\discriminateSet$.

\noindent\textbf{Single-circuit encoding.}\quad
Equation~\eqref{equ:singlecirc} combines the one-hot constraints, the signal constraints, and the aggregated gate-semantics constraints into a single-circuit CNF encoding for the public topology~$\Topology$. Any satisfying assignment yields a concrete gate-type assignment consistent with all observations in~$\discriminateSet$.

\noindent\textbf{Two-circuit encoding.}\quad
Equation~\eqref{equ:twocirc} instantiates two disjoint copies of the single-circuit encoding under the same topology. For the $j$-th circuit copy, $\VariableGateAssignment_j$ denotes the corresponding duplicated selector-variable family. This encoding is used to reason jointly about two candidate gate assignments that are both consistent with~$\discriminateSet$.

\noindent\textbf{Discriminating-input encoding.}\quad
Equation~\eqref{equ:finddis} searches for a primary-input vector $X$ under which the two circuit copies produce different primary outputs. In this encoding, the superscript $(X)$ labels the additional signal variables induced by the candidate discriminating input $X$, and is distinct from the sample index $i$ used for observed pairs in~$DI$. The formula enforces that both circuit copies are evaluated under the same input $X$, and requires that the two resulting output vectors differ on at least one primary output bit. This primitive is used to generate new oracle queries in the iterative search procedure.

\noindent\textbf{Blocking concrete assignments.}\quad
Equation~\eqref{equ:block} excludes a concrete gate-type assignment $\gateAssignment$ from future satisfying assignments. The clause enforces that the symbolic assignment encoded by $\VariableGateAssignment$ must differ from $\gateAssignment$ on at least one gate.

\medskip
\noindent\textit{Optional admissible-set restriction.}\quad
In all primitives above, the admissible gate-type set $\Allowed(g)$ of each gate $g$ may be restricted by the topology-based preprocessing of Section~\ref{sec:simplificationtheorems}.

\section{Baseline Attack}\label{sec:baseline}

\begin{algorithm}[!tb]
\caption{Baseline SAT-based Function Recovery}
\label{alg:baseline}
\begin{algorithmic}[1]
\Require Circuit topology $\Topology$, oracle $\Oracle{\cdot}$
\Ensure A gate-type assignment $\gateAssignment$ whose instantiation on $\Topology$ is functionally equivalent to the target circuit
\State $\discriminateSet \gets \emptyset$
\While{true}
    \State Find $\gateAssignment_1,\gateAssignment_2,\inputvector$ such that $\gateAssignment_1,\gateAssignment_2$ satisfy $\discriminateSet$ and differ on $\inputvector$
    \If{such $(\gateAssignment_1,\gateAssignment_2,\inputvector)$ exists}
        \State $\outputvector \gets \Oracle{\inputvector}$
        \State $\discriminateSet \gets \discriminateSet \cup \{(\inputvector,\outputvector)\}$
    \Else
        \State \textbf{break}
    \EndIf
\EndWhile
\State Solve for $\gateAssignment$ satisfying $\discriminateSet$
\State \Return $\gateAssignment$
\end{algorithmic}
\end{algorithm}

In this section, we establish a SAT-based baseline attack for Threat Model A by adapting the decamouflaging framework of El Massad et al. \cite{el2015integrated}. \fixed{\reviewnotemulti{\ref{ra:improvement}, \ref{rb:weaknesses}, \ref{ae:theoretical:scalability}} In their original setting, the focus is on scenarios in which only a small subset of gates is camouflaged, and the attacker leverages a discriminating input set together with SAT solving to pinpoint those hidden positions. In contrast, our GH threat model conceals the type of every gate behind uniform encodings, substantially enlarging the search space.} We generalize the same counterexample-guided strategy to this more demanding scenario. 

The idea involves constructing a discriminating I/O pair set $\discriminateSet = $ $ \{(\inputvector, \outputvector)\}$, where each pair ($\inputvector, \outputvector$) consists of an input vector and the corresponding output obtained by oracle query $\Oracle{\inputvector}$. This set is iteratively expanded to eliminate incorrect candidate gate-type assignments.

\fixed{\reviewnotemulti{\ref{rb:algorithm1},\ref{rb:algorithm:explanation},\ref{ae:7}}Algorithm~\ref{alg:baseline} summarizes the baseline SAT-based recovery procedure. Starting from an empty discriminating set $\discriminateSet$, the solver searches for two gate-type assignments $\gateAssignment_1$ and $\gateAssignment_2$ that are both consistent with all I/O pairs currently in $\discriminateSet$, together with an input vector $\inputvector$ on which the two candidate assignments produce different output vectors. If such a triple exists, the oracle $\Oracle{\cdot}$ is queried on $\inputvector$ to obtain the corresponding output $\outputvector$. The new observation $(\inputvector,\outputvector)$ is then added to $\discriminateSet$, thereby excluding all gate-type assignments inconsistent with this oracle response. In our implementation, we additionally block previously found gate-type assignments that become inconsistent with the accumulated observations, which improves solving performance in subsequent iterations. This process continues until no further discriminating input can be found. At that point, $\discriminateSet$ is complete, in the sense that any two gate-type assignments consistent with all pairs in $\discriminateSet$ induce circuits that are functionally indistinguishable over the full input space. Finally, one satisfying gate-type assignment is extracted; when instantiated on the given topology, it defines a circuit functionally equivalent to the target circuit. For a formal proof of completeness and correctness, see~\cite{el2015integrated}.}

The solver calls in the baseline attack are instantiated using the encoding primitives introduced in Section~\ref{subsec:circuitencoding}. In particular, each baseline iteration solves one SAT instance formed by conjoining $\EncodeTwoCircTopo(\Topology,\VariableGateAssignment_1,\VariableGateAssignment_2,\discriminateSet)$ with $\FindDisInputTwoCirc(\Topology,\VariableGateAssignment_1,\VariableGateAssignment_2,X)$. A satisfying assignment to this combined formula simultaneously yields two candidate gate-type assignments consistent with the current discriminating set and a discriminating input on which the two instantiated circuits disagree. Each time a new oracle response $(\inputvector,\outputvector)$ is obtained, the discriminating set is extended and the SAT instance is refined accordingly, so that all assignments inconsistent with the new observation are automatically excluded in subsequent solver calls.

We implement this iterative procedure using incremental SAT solving: all iterations share a single persistent solver instance, to which constraints induced by newly observed I/O pairs and blocking clauses are added progressively. For a given circuit, the complete discriminating set obtained by the procedure is not necessarily unique. Its size depends on the particular sequence of discriminating inputs found during solving, which in turn may vary with the solver's internal search behavior. Accordingly, the size of a complete discriminating set for a fixed circuit is not our primary focus in this work.

\begin{algorithm}[!tb]
\caption{Optimized SAT-based Function Recovery under Threat Model~A}
\label{alg:OptimizedAttackIntuition}
\begin{algorithmic}[1]
\Require Circuit topology $\tau$, oracle $\Oracle{\cdot}$
\Ensure A gate-type assignment $T$ such that its instantiation on $\tau$ is functionally equivalent to the target circuit
\State Initialize the admissible gate-type domains using the Simplification Theorems
\State $DI \gets \emptyset$
\While{true}
    \State Find a reference assignment $T_1$ consistent with $DI$
    \While{there exists an alternative assignment $T_2 \neq T_1$ consistent with $DI$}
        \State Find such an assignment $T_2$
        \State Find a discriminating input $\mathbf{x}$ between $T_1$ and $T_2$
        \If{such an input $\mathbf{x}$ exists}
            \State $\mathbf{z} \gets \Oracle{\mathbf{x}}$
            \State $DI \gets DI \cup \{(\mathbf{x},\mathbf{z})\}$
            \State Globally block assignments inconsistent with $(\mathbf{x},\mathbf{z})$
            \State \textbf{break} to the next outer-loop iteration
        \Else
            \State Temporarily block $T_2$ in the current enumeration
        \EndIf
    \EndWhile
    \State \Return $T_1$
\EndWhile
\end{algorithmic}
\end{algorithm}

\section{Our Optimized Attack}
\label{sec:advancedAttack}

We present an optimized SAT-based function-recovery attack under Threat Model~A. In contrast to the baseline attack, which jointly searches for two candidate gate-type assignments and a discriminating input within a single SAT instance, the optimized attack decomposes the recovery procedure into a sequence of smaller SAT queries that separately compute a reference assignment, enumerate alternative assignments, and search for a distinguishing input for each candidate pair.

The solver first computes a reference gate-type assignment $\gateAssignment_1$ consistent with all currently observed I/O pairs in $\discriminateSet$. It then searches for another candidate assignment $\gateAssignment_2$ that is also consistent with $\discriminateSet$, and uses a separate SAT query to determine whether there exists an input that distinguishes $\gateAssignment_2$ from the reference assignment. If such a discriminating input is found, the oracle is queried on that input, and the resulting observation is added to $\discriminateSet$. The subsequent SAT queries are then solved under the refined constraint set, which excludes all gate-type assignments inconsistent with the new observation. If no discriminating input exists for the current pair $(\gateAssignment_1,\gateAssignment_2)$, the attack blocks $\gateAssignment_2$ and continues searching for another candidate consistent with $\discriminateSet$.

The procedure terminates when, for the current reference assignment $\gateAssignment_1$, no additional candidate gate-type assignment consistent with $\discriminateSet$ remains. At that point, the discriminating set is complete, and $\gateAssignment_1$, instantiated on the given topology, yields a circuit functionally equivalent to the target circuit.


The optimized attack can further incorporate the topology-preserving simplification theorems from Section~\ref{sec:simplificationtheorems} as a preprocessing step. These simplifications restrict the admissible gate-type set of each gate using only the public circuit topology, thereby reducing the candidate search space before SAT solving begins. Algorithm~\ref{alg:OptimizedAttackIntuition} summarizes the optimized SAT-based attack procedure with this preprocessing step enabled. The same optimized attack framework extends to Threat Model~B; the corresponding attack is presented in Section~\ref{sec: extended attack}. In our implementation, the resulting sequence of SAT queries is solved incrementally. Permanent constraints are retained in persistent solver instances, while query-specific temporary constraints are supplied as assumptions and therefore affect only the current SAT call.


The solver calls in the optimized attack are instantiated using the encoding primitives introduced in Section~\ref{subsec:circuitencoding}. Both reference-assignment generation and alternative-assignment enumeration are realized by solving $\EncodeCircTopo(\Topology,\VariableGateAssignment,\discriminateSet)$ together with blocking clauses of the form $\BlockGateAssignment(\VariableGateAssignment,\gateAssignment)$, which exclude previously explored concrete assignments. Given two concrete candidate assignments, discriminating-input search is carried out by a separate SAT query based on $\EncodeTwoCircTopo$ and $\FindDisInputTwoCirc$.



A potential concern is that the inner enumeration loop may become inefficient when many candidate gate-type assignments satisfy the current discriminating set yet remain functionally indistinguishable under the given topology. To mitigate this, we cap the enumeration at $N_{\max}$ candidates and then perform a monolithic SAT check, as in the baseline attack, with the reference assignment fixed to $\gateAssignment_1$. This check determines whether there still exists another satisfying assignment that can be distinguished from $\gateAssignment_1$. If the query is UNSAT, the enumeration phase terminates soundly. In all experiments, we use $N_{\max}=3$, which was sufficient on our benchmark set to avoid excessive inner-loop iterations.

\section{Extended Attack under Hidden Inputs}
\label{sec: extended attack}

We now extend our optimized attack framework to Threat Model~B, where only a subset of the circuit inputs is controllable by the adversary. In this setting, the adversary queries the oracle on attacker-controlled inputs $\inputvector$ and observes outputs $\evalcircuit{\originalCircuit}(\inputvector,\fixedinput)$, where $\fixedinput$ is unknown but fixed throughout the attack. Thus, the adversary observes the behavior of the target circuit only over the attacker-controlled input domain, rather than over the full input space. Accordingly, the recovery task is to find both a gate-type assignment $\gateAssignment$ and a hidden-input assignment $\fixedinput$ such that, when instantiated together on the given topology, they yield a circuit whose behavior matches that of the target circuit over the attacker-controlled input domain.

To this end, we extend the SAT-based recovery procedure of Threat Model~A by augmenting the encoding with Boolean variables for the bits of $\fixedinput$, so that the solver searches the joint space $|\mathcal{L}|^{k}\times\{0,1\}^{n'}$, where $|\mathcal{L}|^{k}$ denotes the space of gate-type assignments for the $k$ hidden gates and $\{0,1\}^{n'}$ denotes the space of assignments to the $n'$ hidden input bits. All topology-preserving simplification theorems remain applicable, and the resulting restrictions on each gate's admissible gate-type set are applied before each SAT invocation to reduce the combined search space. In each iteration, the solver proceeds as follows:

\begin{enumerate}
    \item \textbf{Candidate generation:} Solve the SAT instance to obtain a candidate pair $(\gateAssignment_1, Y_1)$ that satisfies all constraints induced by the current discriminating set $\discriminateSet$.
    \item \textbf{Enumeration of alternatives:} Add a blocking clause that excludes the current joint assignment $(\gateAssignment_{1}, Y_{1})$, and invoke the solver again to obtain another satisfying pair $(\gateAssignment_{2}, Y_{2})$.
    \item \textbf{Discriminating input search:} Fix both candidate pairs $(\gateAssignment_1, Y_1)$ and $(\gateAssignment_2, Y_2)$ and solve for an input vector $\inputvector\in\{0,1\}^n$ such that $\evalcircuit{(\Topology,\gateAssignment_1)}(\inputvector, Y_1) \neq \evalcircuit{(\Topology,\gateAssignment_2)}$ $(\inputvector, Y_2)$.
    If such an $\inputvector$ exists, the corresponding oracle output is queried and the resulting pair $(\inputvector, \outputvector)$ is added to $\discriminateSet$.
\end{enumerate}

The discriminating set $DI$ is complete once no additional candidate pair satisfies the accumulated constraints and is distinguishable from $(T_1,Y_1)$ by any attacker-controlled input. At that point, all remaining candidate pairs are indistinguishable over the attacker-controlled input domain. A final SAT query then returns one such pair $(\gateAssignment, Y)$, which, when instantiated on the given topology, yields a circuit functionally equivalent to the target circuit over that domain.

\section{Benchmark Circuits}\label{sec:benchmarks}

\begin{figure}[!t]
    \centering
    \includegraphics[width=0.9\linewidth]{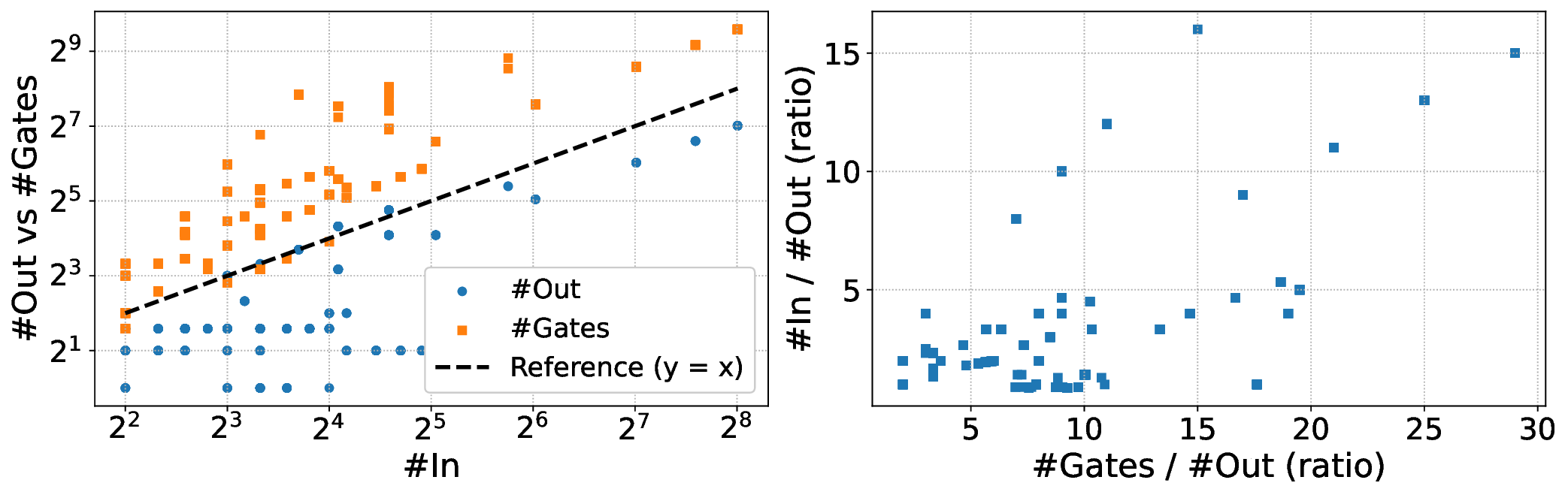}     
    \caption{Structural characteristics of the evaluated benchmark circuits. Left: numbers of outputs and gates versus the number of primary inputs on a log--log scale, with a \(y=x\) reference line. Right: distribution of circuits by gates-to-outputs and inputs-to-outputs ratios.}
    \label{fig:circuitsize}
\end{figure}

All evaluated circuits are combinational Boolean circuits, consistent with the model assumed in garbled-circuit protocols. This restriction isolates the function-recovery problem induced by gate hiding from orthogonal challenges such as sequential state reconstruction. Point functions, representing an extreme case for function recovery, are analyzed separately in Section~\ref{sec:pointfunctions}.



\fixed{\reviewnotemulti{\ref{ae:theoretical:scalability}}Our benchmark set comprises three classes of circuits, described below. Fig.~\ref{fig:circuitsize} summarizes the distribution of circuit sizes over the full evaluation suite. Excluding point functions, the set contains 72 circuits of varying sizes, with 2 to 257 primary inputs, 2 to 129 outputs, and 4 to 768 gates. For the 2PC and MPC benchmarks, the circuits are synthesized using Synopsys Design Compiler over the full library of sixteen two-input Boolean gates.}



\begin{enumerate}
    \item \textbf{ISCAS Benchmarks.}
    We include standard benchmarks from ISCAS'89. Following \cite{el2015integrated}, sequential circuits are converted to combinational form by unrolling one cycle and replacing each flip-flop with a primary input. Gates with fan-in greater than two are decomposed into cascaded two-input gates.

    \item \textbf{2PC Circuits.}
    We evaluate three families of commonly used two-party computation circuits: adders, comparators, and Hamming-distance circuits. Each family is instantiated at multiple input sizes to assess scalability and structural variation. \fixed{\reviewnotemulti{\ref{ra:improvement}} These circuits were not considered by \cite{el2015integrated}.}

    \item \textbf{\fixed{\reviewnotemulti{\ref{rb:weaknesses},\ref{rc:testsuite},\ref{rc:2},\ref{ae:3}}MPC Circuits.}}
    To assess MPC-specific logic, we include compact combinational circuits derived from the privacy-preserving sensor-fusion algorithms of \cite{jin2024pg}. The number of sensors ranges from 3 to 19. Although modest in absolute size, these circuits preserve the algorithmic structure and data-flow dependencies of the original designs.
\end{enumerate}




\textbf{Comparison with Decamouflaging Benchmarks.}
\fixed{\reviewnotemulti{\ref{ra:improvement}, \ref{rb:weaknesses}}Prior SAT-based decamouflaging studies typically considered partial camouflage, in which only a small fraction of gates is hidden.
For example, El Massad et al.~\cite{el2015integrated} followed the setting of \cite{rajendran2013security}, camouflaging fewer than 10\% of the gates in large ISCAS circuits and modeling each camouflaged gate with two or three possible types.
Their largest evaluated configuration corresponds to a search space of approximately $3^{350}$.
In contrast, our benchmarks include circuits with up to 768 gates, in which every gate may realize any of sixteen two-input Boolean functions, yielding an effective search space on the order of $16^{768}$.} \fixed{These observations suggest that gate count alone is not a sufficient indicator of recovery difficulty across these settings. Even moderate-sized circuits under our gate-hiding model can induce substantially larger uncertainty than much larger circuits subject to partial camouflage.}

\section{Experimental Evaluation}\label{sec:implementation}




We evaluate SAT-based function-recovery attacks under two threat models, denoted Model~A and Model~B. For each model, we compare the baseline and optimized attacks, and for each attack we further evaluate performance with and without the proposed simplification theorems. This experimental design separates the effect of the optimized attack procedure from that of the simplification-based preprocessing.



All attacks are executed on AWS EC2 c7i.2xlarge instances (8 vCPUs, 16 GiB RAM), with each run constrained by both a 24-hour time budget and the fixed compute resources of the assigned instance. We report only circuits that are successfully recovered within these limits. For a subset of unsuccessful cases, we additionally repeated the attack to assess whether the observed failure was primarily due to the particular oracle-query sequence. The implementation is written in Python~3 and uses Glucose3 via PySAT for SAT solving.

\subsection{Experimental Results} \label{sec:results}

\begin{table*}[t]
\begin{adjustbox}{max width=\textwidth}
\begin{threeparttable}
\begin{tabular}{lcccccllll}
\toprule
Circuit & \#In & \#Out & \#Gates & \#S & \#Z & Opt+Simpl & Opt & Baseline+Simpl & Baseline \\
\midrule

s27   & 7  & 3  & 8   & 0  & 0  & ZSR, 0.12 (14)       & 0.30 (12)        & ZSR, 0.21 (14)       & 0.45 (15) \\
s27a  & 7  & 3  & 8   & 0  & 0  & R, 0.08 (10)         & 0.33 (12)        & R, 0.20 (15)         & 0.63 (16) \\
s298  & 17 & 20 & 125 & 14 & 10 & R, 329.17 (65)       & 1,484.13 (69)    & ZS, 629.52 (107)     & 2,211.67 (102) \\
s344  & 24 & 17 & 109 & 1  & 8  & ZS, 344.84 (35)      & 1,805.68 (46)    & ZSR, 420.04 (54)     & 6,330.31 (60) \\
s349  & 24 & 17 & 112 & 1  & 8  & ZS, 251.27 (48)      & 1,545.86 (41)    & ZSR, 531.56 (46)     & 8,236.99 (43) \\
s382  & 24 & 27 & 148 & 8  & 25 & ZSR, 816.39 (71)     & 3,299.32 (65)    & ZSR, 988.87 (85)     & 5,207.41 (80) \\
s386  & 13 & 13 & 188 & 6  & 5  & ZS, 547.62 (197)     & 11,262.23 (188)  & ZSR, 1,047.75 (208)  & 15,934.71 (190) \\
s400  & 24 & 27 & 158 & 9  & 27 & R, 1,724.17 (66)     & 3,328.19 (60)    & R, 2,865.13 (75)     & 5,748.88 (83) \\
s444  & 24 & 27 & 171 & 5  & 22 & R, 1,179.96 (66)     & 4,962.10 (60)    & R, 1,767.53 (86)     & 12,316.01 (98) \\

\bottomrule
\end{tabular}
\begin{tablenotes}
\footnotesize
\item[$^*$] The attack runtime is reported in seconds, and the size of the discriminating set is shown in parentheses. ``-'' indicates time-out.
\end{tablenotes}
\end{threeparttable}
\end{adjustbox}
\caption{Performance Comparison on ISCAS89 Benchmarks under Model A.}
\label{tab:standardcircuits}
\end{table*}

\textbf{ISCAS'89 benchmark.} To facilitate comparison with prior work, we first evaluate our optimized attack on 
selected ISCAS'89 benchmark circuits under Model~A. The results are 
summarized in Table~\ref{tab:standardcircuits}. For each circuit, the table 
reports its number of primary inputs and outputs, total gate count 
(excluding $\gate{NOT}$ gates), and the number of S-Class and Z-Class 
gates. 
Each 
entry lists the recovery runtime in seconds, followed by the size of the 
resulting discriminating set in parentheses (number of oracle queries). Across the ISCAS'89 benchmark circuits, the optimized approach significantly outperforms 
the baseline. The improvement is most pronounced when the 
simplification theorems are enabled. For example, on circuit s349, the optimized attack with ZS simplification completes function recovery in 251.27 s, whereas the baseline requires 8236.99 s, yielding a speedup of approximately 32.78x. The two approaches recover complete discriminating input sets of sizes 48 and 43, respectively.

\begin{figure*}[t]
    \centering
    \includegraphics[width=1\linewidth]{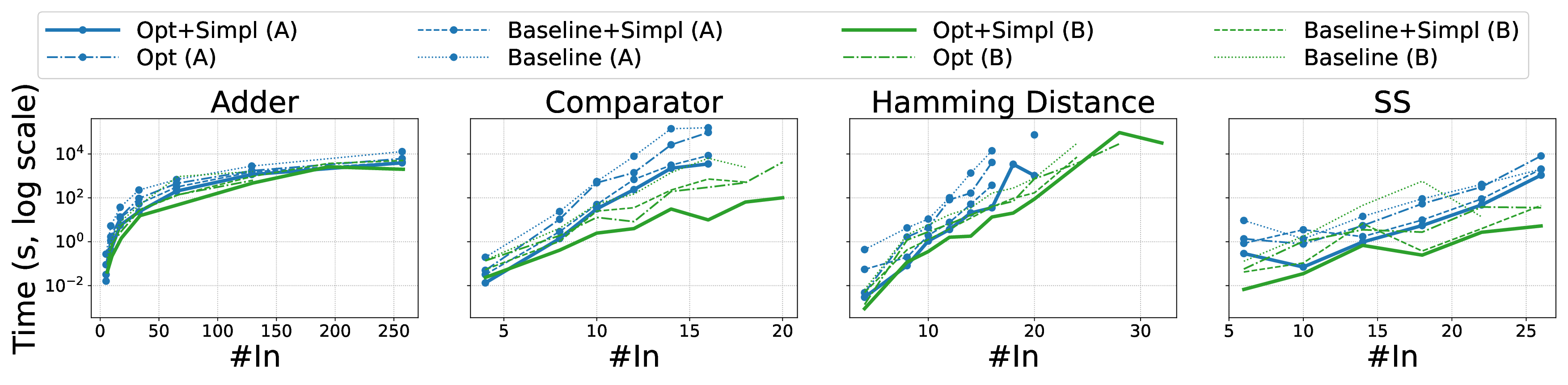}
    \caption{Log-scale recovery runtime on representative MPC functions.}
    \label{fig:runtimeinput}
\end{figure*}

\begin{figure*}
    \centering
    \includegraphics[width=0.8\linewidth]{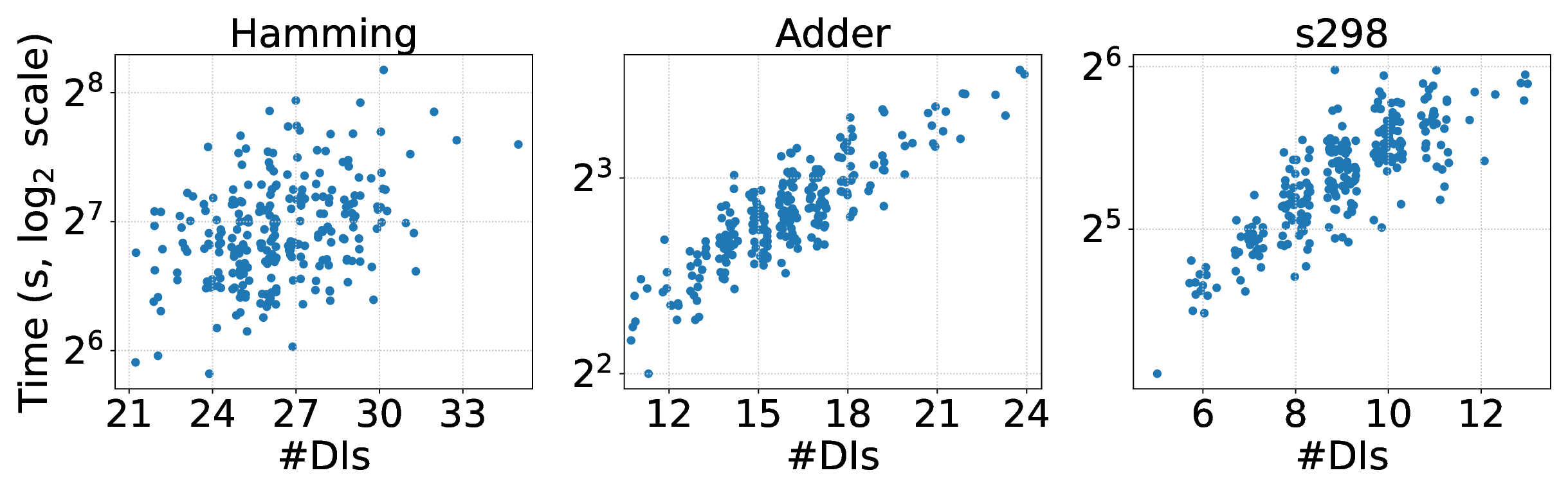}
    \caption{Exhaustive evaluation of Model~B over \(2^8=256\) fixed-input assignments. }
    \label{fig:modelBexhasutive}
\end{figure*}

\noindent\textbf{2PC and MPC circuits.} Fig.~\ref{fig:runtimeinput} reports the log-scale recovery time for Adder, Comparator, Hamming Distance, and SS (the Schmid-Schossmaier interval-fusion algorithm used in fault-tolerant sensor fusion~\cite{jin2024pg}), under both Model~A and Model~B. For the experiments under Model B, half of the circuit inputs are designated as hidden and fixed to randomly selected values. At the scale of Fig.~\ref{fig:runtimeinput}, the randomness of these hidden inputs does not appear to materially alter the overall recovery trend. We examine assignment-specific variation separately in Fig.~\ref{fig:modelBexhasutive}. Each subplot corresponds to a fixed functionality instantiated with different input sizes. All functions are evaluated under the same four attack configurations used in Table~\ref{tab:standardcircuits}. The solid green and solid blue lines in Fig.~\ref{fig:runtimeinput} denote the recovery times of the optimized and baseline attack procedure, respectively, both with simplification preprocessing. In Model~B, the hidden fixed input vector $\fixedinput$ reduces the effective input domain from $n+n'$ bits to $n$ bits, which helps explain why the results for Model~B are often smaller than those for Model~A. For several larger instances, missing data points indicate that the corresponding algorithm either failed to finish within the 24-hour time budget or ran out of memory on the AWS instance due to the growth of the discriminating set and the resulting increase in SAT constraints. Although the fully optimized attack (Opt+Simpl) does not always achieve the lowest recovery time for small circuits, the performance gap between different approaches widens with increasing circuit size, and Opt+Simpl consistently becomes the most efficient strategy for larger instances. \fixed{\reviewnotemulti{\ref{ra:weaknesses}\ref{ra:adder}}Among the four functions, the Adder benchmark shows the most regular trend across input sizes, and the gap between solver configurations is smaller than for the other functions. This behavior appears to be benchmark-specific, likely due to the regular structure and simple functionality of adders rather than a general scaling trend. Because the figure is plotted on a logarithmic scale, constant-factor improvements are also visually compressed.}

\noindent\textbf{Fixed-Input Effects in Model~B.} \fixed{\reviewnotemulti{\ref{rb:weaknesses},\ref{rd:2},\ref{rd:modelb},\ref{ae:notpoint},\ref{ae:8}}To further characterize the impact of fixed inputs in Model~B, we exhaustively evaluated three circuits: Adder, Hamming Distance, and s298. In each case, the hidden fixed portion was set to half of the inputs, rounded down, giving 8 fixed bits for Adder and Hamming Distance (16 inputs each) and also 8 fixed bits for s298 ($\lfloor 17/2 \rfloor$). Thus, each circuit induces $2^8 = 256$ fixed-input assignments. For each assignment, we ran the recovery procedure and recorded both the number of oracle queries (\#DIs) and the runtime. As shown in Fig.~\ref{fig:modelBexhasutive}, recovery cost varies across fixed-input assignments even for the same circuit, although most cases remain concentrated in a central region of the distribution. In general, larger \#DI counts are associated with longer runtimes, with a clearer trend for Adder and s298 than for Hamming Distance. Overall, these results show that, in Model~B, the recovery cost is sensitive to the fixed-input assignment, and that this sensitivity varies across circuits.
}

\begin{figure}[t]
    \centering
    \includegraphics[width=0.7\linewidth]{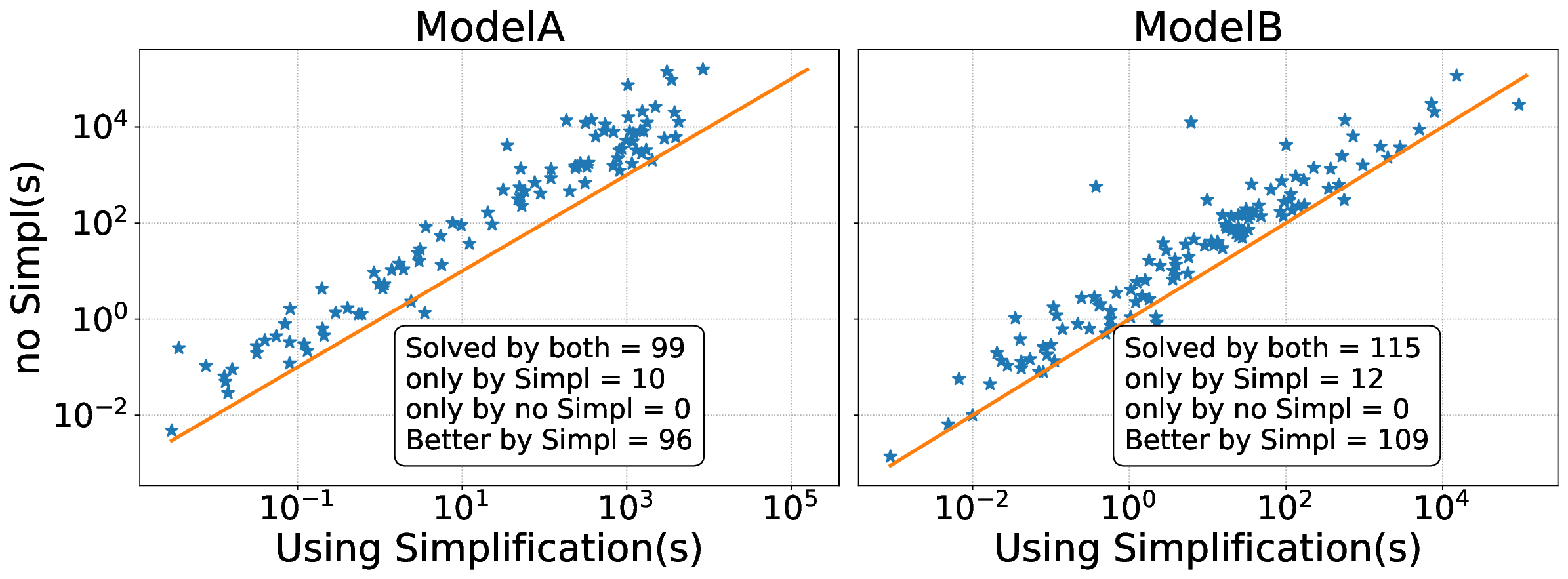}
    \caption{Overall Impact of Simplifications}
    \label{fig:impact:simplification}
\end{figure}

\begin{figure}[t]
    \centering
\includegraphics[width=0.7\linewidth]{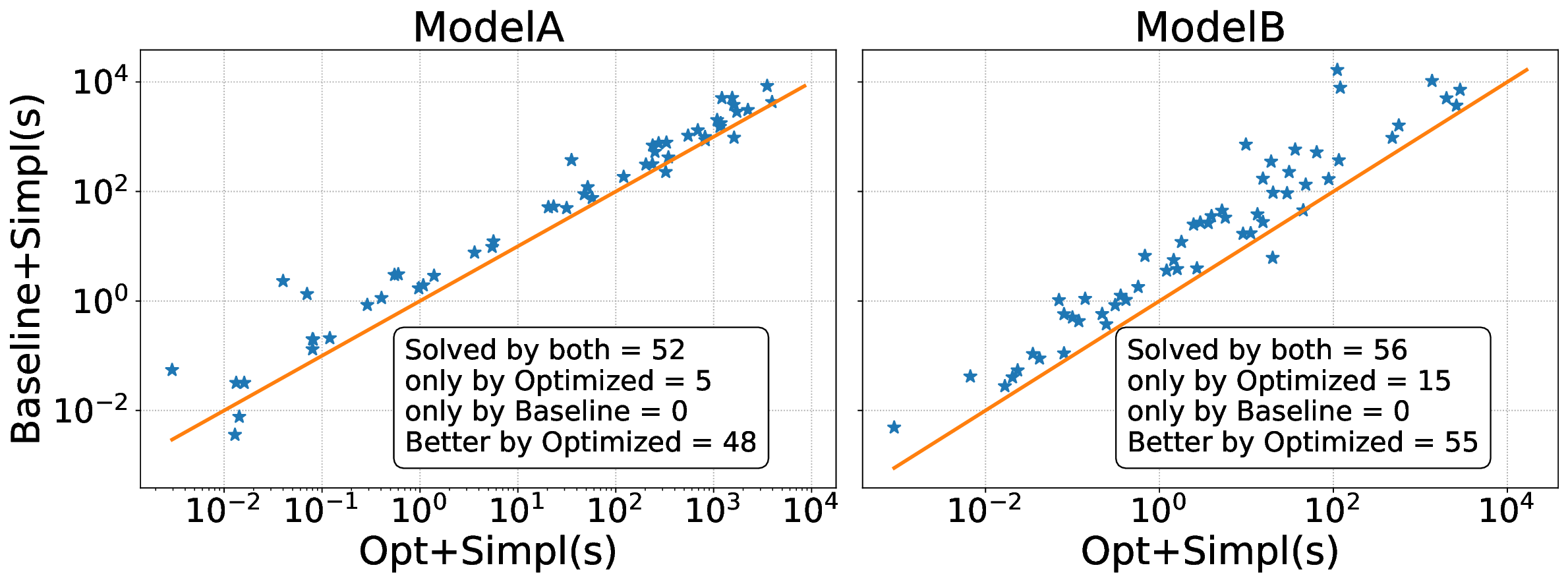}
    \caption{Impact of Baseline vs Optimized Attack Procedure} \label{fig:impact:algorithm}
\end{figure}


\noindent\textbf{With vs.\ Without Simplification Preprocessing.} Fig.~\ref{fig:impact:simplification} isolates the effect of gate-type simplification by comparing, for each benchmark circuit, the recovery runtime with simplification (x-axis) against the runtime without simplification (y-axis), under Models A and B. Each circuit contributes up to two points, one for the baseline and one for our optimized attack, so the plot evaluates the simplification step consistently across algorithms. Points above the diagonal indicate a speedup from simplification, while points below indicate a slowdown. Cases that succeed only with simplification have no corresponding no-simplification runtime and are therefore omitted; their counts are reported in the in-plot annotations. Overall, the concentration of points above the diagonal shows that simplification yields broad and consistent runtime reductions in both models.

\noindent\textbf{Optimized vs.\ Baseline Attack.} Fig.~\ref{fig:impact:algorithm} plots the recovery time of the two algorithms on the same set of instances in Model~A and Model~B, where each point corresponds to one circuit instance. The diagonal line indicates equal runtime for both approaches. Points below the diagonal therefore represent instances where the optimized attack is slower. In both threat models, the majority of instances lie above the diagonal, indicating that the optimized attack consistently achieves a lower recovery time than the baseline. Moreover, many points deviate substantially from the diagonal, reflecting speedups of multiple orders of magnitude on a logarithmic scale. The summary statistics further confirm this trend: for most instances solved by both algorithms, the optimized attack is faster, and it additionally succeeds in several instances where the baseline fails to terminate within the time limit.

In summary, our experiments show that the proposed optimized SAT attack reduces recovery time compared with the baseline on most tested circuits, with additional gains from topology-preserving simplification preprocessing. These improvements are observed under both threat models, while the number of oracle queries remains far smaller than the full input space of the target circuit.

\subsubsection{Circuit Complexity and Discriminating Inputs} \label{sec:circuitcomplecity}

\begin{figure*}
    \centering
    \includegraphics[width=\linewidth]{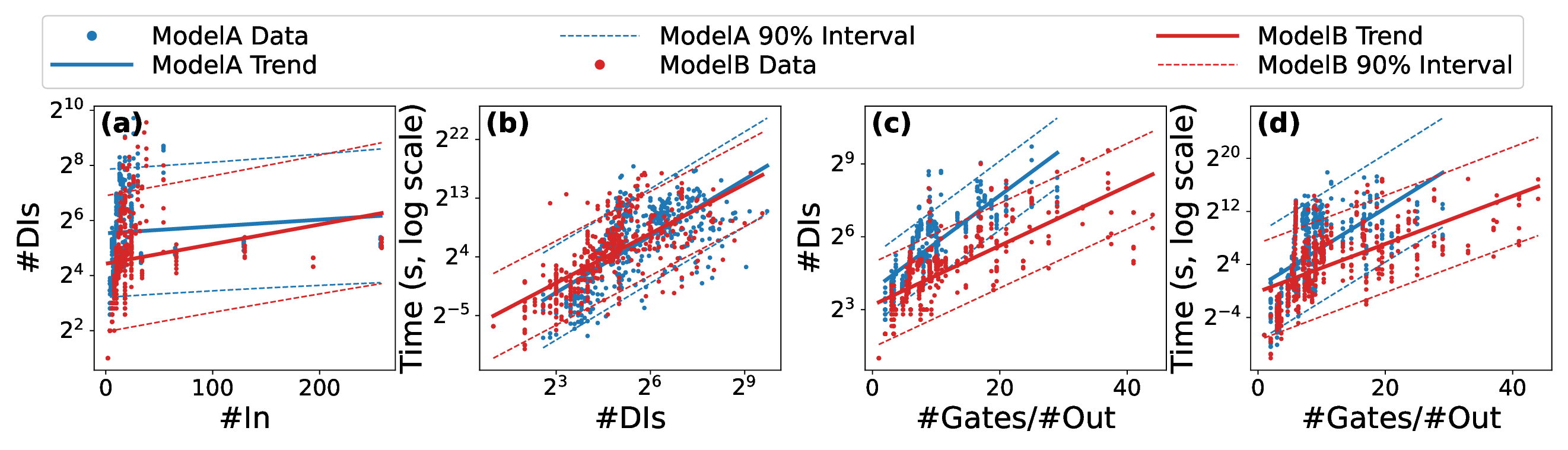}
    \caption{Relationships between discriminating inputs (DIs), recovery time, and circuit structural metrics for Models A and B. }
    \label{fig:DIs-Circuit-Size}
\end{figure*}
\fixed{\reviewnotemulti{\ref{ra:question},\ref{rb:weaknesses},\ref{rd:5},\ref{rd:8}}
Since discriminating inputs (DIs) are used to separate candidate gate-type assignments, the required \#DI count serves as an informative indicator of recovery difficulty. We next examine how this quantity relates to circuit characteristics. Fig.~\ref{fig:DIs-Circuit-Size} summarizes the relationships between \#DI count, recovery time, and circuit structure for all solved instances in our main benchmark set, excluding point functions. 

As shown in Fig.~\ref{fig:DIs-Circuit-Size}(a), the number of primary inputs alone is not a reliable indicator of recovery difficulty. The weak correlation between the number of primary inputs and the number of discriminating inputs required suggests that function recovery is not driven primarily by the size of the input space.

In contrast, Fig.~\ref{fig:DIs-Circuit-Size}(c) shows that \#DI exhibits a clear increasing trend with respect to the ratio between the number of gates and outputs.
This ratio captures the amount of internal logic that must be inferred per observable output and therefore reflects the structural complexity of the circuit.
Across the evaluated benchmarks, the $\#\text{Gates}/\#\text{Out}$ ratio ranges from 2 to 45.
Circuits with larger ratios consistently require more discriminating inputs. Fig.~\ref{fig:DIs-Circuit-Size}(c, d) show that both solver runtime and the number of discriminating inputs increase with this structural complexity metric.
This indicates that the $\#\text{Gates}/\#\text{Out}$ ratio is a useful predictor for both oracle-query complexity and SAT-solving time.

The number of discriminating inputs (\#DI) required for complete recovery varies across benchmarks, ranging from 4 to 586 (median 20, mean 38.5). Nevertheless, for most benchmarks, the required \#DI count remains far smaller than the corresponding full input space. In addition, for a given circuit, the \#DI counts under the four attack configurations (i.e., Baseline, Baseline+Simpl, Opt, Opt+Simpl) are broadly similar and remain within the same order of magnitude, suggesting that the observed runtime improvements arise primarily from easier SAT solving rather than from substantially fewer oracle queries.
}

\fixed{\reviewnotemulti{\ref{rc:3},\ref{ae:3}}}\section{Testing the Limits: Point Functions}
\label{sec:pointfunctions}

\begin{figure}
    \centering
\includegraphics[width=\linewidth]{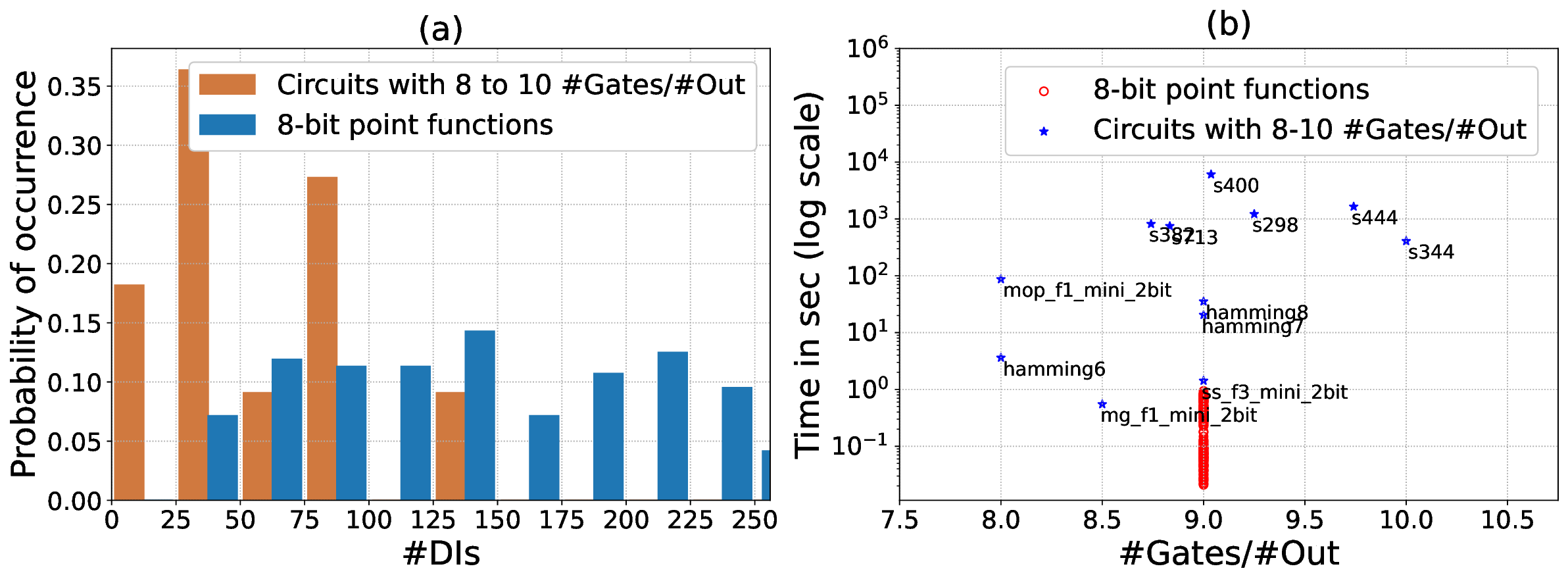}
    \caption{8-bits Point Functions vs Circuits with similar \#Gates/\#Out ratio}
    \label{fig:pointfunctions}
\end{figure}
\fixed{\reviewnotemulti{\ref{ae:notpoint},\ref{ae:9},\ref{ac:pointfunction},\ref{rc:3},\ref{ae:3}}
While the preceding experiments characterize recovery behavior on a broad benchmark set, point functions probe a qualitatively different and more extreme regime. They therefore provide a useful stress test for the limits of SAT-based function recovery. Over an $n$-bit input space, a point function outputs 1 on exactly one input and 0 on all others. In our attack setting, however, the adversary is not told in advance that the target circuit implements a point function, so this structure must be inferred from observations rather than assumed.

We begin with an attacker who has only black-box oracle access and no circuit-topology side information. For a target function with input size $n$, the space of possible Boolean functionalities has size \(2^{2^{n}}\). In this setting, the attacker cannot certify that the oracle implements a point function without querying the entire input space, because any unqueried input may output 1. 

By contrast, when the circuit topology is known, it constrains the set of realizable functionalities: some Boolean functions cannot be represented by the given topology under any gate-type assignment. The attacker can therefore rule out part of the space of candidate functionalities even without any oracle query. The topology then consistently constrains the space of candidate functionalities in conjunction with the oracle I/O behaviors observed by the attacker. Therefore, it can prune inconsistent global assignments without enumerating all $2^{n}$ inputs and substantially reduce best-case and average-case recovery cost.

However, the circuit topology does not identify which input corresponds to the unique output-1 point, since different point functions over the same input domain can be realized using the same topology. Therefore, even when the topology is known, exact recovery in the worst case may still require querying the entire input space.

To study this boundary case, we exhaustively evaluated all 8-bit point functions
and compared them with circuits having a similar $\#\text{Gates}/\#\text{Out}$
ratio. Fig.~\ref{fig:pointfunctions}(a) shows that point
functions may require more discriminating inputs than other circuits of
comparable structural complexity, Fig.~\ref{fig:pointfunctions}(b) shows
that recovery time is often smaller for point functions due to their concise
circuit structure. For each target function, we measured the number of oracle queries required for exact recovery under the given topology. The required number of queries ranged from 29 to 252, with an average of 136.56 and a median of 131. Relative to the full input space of size \(256\), this corresponds to an average reduction of about 46.7\% in the number of required oracle queries. These results indicate that, although topology knowledge does not improve the worst-case query complexity of point-function recovery, it can reduce query cost in typical cases by eliminating structurally inconsistent candidates.  
}

\section{Related Work}\label{sec:related_work}
\noindent\textbf{SAT-based decamouflaging.} 
Classical SAT-based decamouflaging assumes that the attacker knows the circuit topology, treats hidden gate types as symbolic variables, and uses oracle-guided discriminating inputs to iteratively eliminate inconsistent assignments until the remaining solutions are functionally equivalent or unique \cite{el2015integrated,liu2016oracle,yu2017incremental}. These works establish the algorithmic foundation most closely related to our setting, and subsequent extensions show that the same paradigm can be pushed to sequential circuits without scan access and analyzed more systematically from the standpoint of attack power and limits of camouflage \cite{el2017reverse,el2019sat}. \fixed{\reviewnotemulti{\ref{ra:weaknesses},\ref{rb:9}}However, prior SAT-based attacks primarily characterize recovery as repeated consistency checking against oracle observations, with scalability addressed mainly through improved formulations such as incremental solving. In contrast, our setting asks how, for circuits with known topology but gate types hidden over the full 16-function Boolean gate set, structural information can be exploited to shrink the candidate gate-type assignment space, and how the recovery task can be organized into smaller SAT queries to improve solver performance without modifying the underlying SAT engine. Furthermore, these works establish how to recover by finding discriminating inputs, but do not explicitly analyze how ambiguity is distributed across gates and how simplification and pruning affect solver complexity when the oracle is weakened, such as under partial input knowledge rather than a standard full-I/O oracle \cite{liu2016oracle,yu2017incremental,el2017reverse}. A related line of work enhances SAT-based recovery with stronger observations, such as side-channel leakage, to prune candidate solutions more aggressively, but such settings rely on information unavailable in our threat model~\cite{shamsi2021circuit}.}

\noindent\textbf{GH Garbled Circuits.}
Semi-private function evaluation relaxes full function privacy by allowing limited structural leakage~\cite{paus2009practical,gunther2019poster,kolesnikov2018free}. Gate-hiding garbled circuits are a prominent instantiation of this design point~\cite{rosulek2017improvements,lin2023skipping,rosulek2021three,wang2024reducing}, where the circuit topology is revealed while individual gate types remain hidden. Prior work in this line focuses primarily on protocol design and efficiency improvements, rather than on the concrete algorithmic exploitability of the revealed topology for function recovery. Our work complements this line by studying that attack surface directly.

Among representative GH constructions, Rosulek~\cite{rosulek2017improvements}
proposed two gate-hiding garbling schemes that leak only circuit topology while
hiding gate types. These schemes improve the supported gate class and rely only
on PRF-based assumptions, but still incur higher per-gate cost than
Free-XOR-enabled non-gate-hiding garbling on XOR-heavy circuits, with one
variant additionally requiring polynomial interpolation. Lin et
al.~\cite{lin2023skipping} later introduced a skipping scheme for GH garbled
circuits based on prime implicants, reducing evaluation work by avoiding
redundant computation paths, at the cost of an additional preprocessing step for
identifying skippable structures. Rosulek and Roy~\cite{rosulek2021three}
further reduced communication by giving a Free-XOR-compatible construction with
$1.5\kappa+O(1)$-bit AND gates and a gate-hiding variant supporting any 2-input
Boolean gate for $1.5\kappa+O(1)$ bits. These works improve the efficiency and
expressiveness of gate-hiding garbling, but do not analyze whether the revealed
topology itself enables practical function recovery.

\noindent\textbf{SAT-based Attacks on Logic Locking.} 
\fixed{\reviewnotemulti{\ref{rb:9}}Logic locking modifies a circuit so that it behaves correctly only under a secret key. SAT-based attacks recover that key by iteratively pruning incorrect candidates using distinguishing input patterns~\cite{subramanyan2015evaluating,shen2018sat,yasin2016sarlock,xie2016mitigating}. Although this line of work shares the use of discriminating inputs and SAT-based pruning, the adversarial objective and uncertainty structure differ fundamentally from the gate-hiding function-recovery setting studied here.}

\section{Limitations of this work}\label{sec:limitations}
We now briefly discuss the main limitations of this work.

\textbf{Minimizing oracle queries.} A given target function may admit multiple complete discriminating-input sets, so exact recovery can in principle be achieved with different sets of oracle queries. Since recovery time is closely related to the number of required discriminating inputs (\#DIs), identifying shorter query sequences is an important problem. However, neither the proposed SAT-based recovery algorithm nor the topology-preserving simplification theorems are designed to minimize the number of oracle queries. Instead, they aim to accelerate recovery for a given search process by reducing the candidate gate-type assignment space and simplifying the SAT instances solved along the way.

\textbf{Scalability.} \fixed{\reviewnotemulti{\ref{ra:baselines},\ref{rb:weaknesses},\ref{rd:3},\ref{ae:theoretical:scalability}}SAT-based function recovery under our gate-hiding setting remains difficult to scale when each hidden gate may take any type from the full gate library. In this case, the candidate gate-type assignment space grows rapidly with circuit size, which in turn increases the difficulty of the SAT instances arising during recovery. Although we propose several techniques to improve attack performance, our experiments still do not scale beyond medium-sized circuits under a 24-hour budget on AWS EC2 c7i.2xlarge instances (8 vCPUs, 16\,GiB RAM). In addition, while parallel and multi-core SAT solvers may offer further improvements, we do not explore them in the current work. We also do not attempt to determine the largest circuit that can be recovered in principle, since sufficiently long runtime and larger compute resources could eventually enable additional cases to complete. Instead, our evaluation is intended to show how much the proposed optimization techniques can accelerate recovery within a fixed practical resource budget.}

\section{Countermeasures and Conclusion}\label{sec:conclusion}

\fixed{\reviewnotemulti{\ref{rc:6}}We conclude by briefly discussing possible countermeasures against our attack on gate-hiding garbled circuits. One direction is to incorporate point-function-like structures that enlarge the required discriminating-input set. Another is to introduce circuit-level redundancy, such as additional gates, expanded input domains, or strategically placed constant gates, in order to enlarge the candidate gate-type assignment space and increase the difficulty of the resulting SAT instances. A further mitigation strategy is to reduce the number of Z-Class and S-Class gates, thereby limiting the effectiveness of our topology-preserving simplifications. More fundamentally, one may avoid exposing the original circuit topology altogether. Universal-circuit-based private function evaluation schemes pursue this goal by hiding topology and therefore mitigate the attack surface exploited here, although they correspond to a different design point from the semi-private gate-hiding setting considered in this work.}

In this work, we show that circuit topology constitutes a meaningful leakage channel in gate-hiding garbled circuits. When the topology is known and gate types are hidden, this structural information can be exploited to accelerate SAT-based function recovery. We further show that topology-preserving simplifications and an optimized SAT-based recovery procedure can substantially reduce recovery time relative to a baseline attack. Overall, these results highlight that circuit topology should be treated as security-relevant leakage when assessing the protection provided by gate-hiding garbled circuits.



\bibliographystyle{alpha}
\bibliography{biblio}

\appendix




\section{Proofs} 
\label{Appendix: proofs}

We first define a set of special type set operators which will facilitate the following proofs of Theorem~\ref{thm:R-wave} and~\ref{thm:sz}, and then we prove the correctness of these two theorems.

\subsection{Type Set Operators}

{\bf Operators on type sets:}
For a type set $\mathcal{T}$ we define the set of gates that negates the output wire of a gate in $\mathcal{T}$ as
$$ \overline{\mathcal{T}}= \{ \mathsf{NOT} \ g \ : \ g\in \mathcal{T}\}.$$
Each gate in $\mathcal{T}$ has two input wires, a left input wire and a right input wire. The following sets extend $\mathcal{T}$ by possibly negating the left and/or right input wire:
\begin{align*}
\leftindex_{\circ}{\underline{\mathcal{T}}}
    &= \mathcal{T} \cup 
       \{\, g(\mathsf{NOT}\ A, B) : g \in \mathcal{T} \,\}, \\[4pt]
\underline{\mathcal{T}}_\circ
    &= \mathcal{T} \cup
       \{\, g(A, \mathsf{NOT}\ B) : g \in \mathcal{T} \,\}, \\[4pt]
\leftindex_{\circ}{\underline{\mathcal{T}}}_\circ
    &= \mathcal{T} \cup
       \{\, g(\mathsf{NOT}\ A, B) : g \in \mathcal{T} \,\} \nonumber \\[2pt]
&\quad \cup
       \{\, g(A, \mathsf{NOT}\ B) : g \in \mathcal{T} \,\} \nonumber \\[2pt]
&\quad \cup
       \{\, g(\mathsf{NOT}\ A, \mathsf{NOT}\ B) : g \in \mathcal{T} \,\}.
\end{align*}

When we write $\mathcal{T}=\mathcal{T}'$ for two type sets $\mathcal{T}$ and $\mathcal{T}'$, then we mean that the two type sets provide gates with {\em equal functionality}. For example,
let ${\Lset}$ be the full set of 16 types. Notice that
$$\overline{{\Lset}} = {\Lset} = \leftindex_{\circ}{\underline{{\Lset}}} = \underline{{\Lset}}_\circ = \leftindex_{\circ}{\underline{{\Lset}}}_\circ $$
where equations relate to functionality. 

The following sets extend $\mathcal{T}$ by possibly replacing the left and/or right input wire by $\gate{TRUE}$:
\begin{align*}
\leftindex_{T}{\underline{\mathcal{T}}}
    &= \mathcal{T} \cup
       \{\, g(\gate{TRUE}, B) : g \in \mathcal{T} \,\}, \\[4pt]
\underline{\mathcal{T}}_T
    &= \mathcal{T} \cup
       \{\, g(A, \gate{TRUE}) : g \in \mathcal{T} \,\}, \\[4pt]
\leftindex_{T}{\underline{\mathcal{T}}}_T
    &= \mathcal{T} \cup
       \{\, g(\gate{TRUE}, B) : g \in \mathcal{T} \,\} \nonumber \\[2pt]
&\quad \cup
       \{\, g(A, \gate{TRUE}) : g \in \mathcal{T} \,\} \nonumber \\[2pt]
&\quad \cup
       \{\, g(\gate{TRUE}, \gate{TRUE}) : g \in \mathcal{T} \,\}.
\end{align*}

\subsection{Proof of Theorem~\ref{thm:R-wave}}

\noindent{\bf R-Wave Simplification:} Let 
\begin{align*}
{\Rset} &= \{\gate{XOR}, \gate{OR}, \mathsf{A\Rightarrow B}, \mathsf{B\Rightarrow A},
            \gate{NAND}, \gate{TRUE}, \mathsf{NOT \ A}, \\
            &\quad \gate{NOT \ B}\} \ \text{and} \\
\overline{\Rset} &= \{\mathsf{XNOR}, \mathsf{NOR}, \mathsf{A \ AND \ (NOT \ B)}, 
                     \mathsf{(NOT \ A) \ AND \ B}, \\
&\quad \gate{AND}, \mathsf{FALSE}, \mathsf{A}, \gate{B}\}.
\end{align*}

Initially, each circuit gate $g$ is in ${\Lset}$ and we have not assigned any new type set to any gate of the circuit. Let $R$ be the set of gates that have been assigned type set ${\Rset}$ -- initially, $R=\emptyset$. We will one by one consider all the gates of the circuit, except those in the primary output layer, and add these to $R$. This procedure will have as invariant that if a gate $g\in R$, then if an input wire of $g$ connects to a gate $g'$, then $g'\in R$ (this is for both input wires of $g$).

As long as there is a gate $g\not\in R$ with each of its input wires either from $R$ or from an actual input to the circuit, and $g$ is not in the primary output layer of the circuit,  we do the following:  Since $g\not\in R$, $g$ still has type set ${\Lset}$ assigned to it.
Since ${\Lset} = {\Rset} \cup \overline{{\Rset}}$, either $g\in {\Rset}$ or $g\in \overline{{\Rset}}$. In the latter case we restrict $g\in {\Rset}$ and push the negation of the output wire to all the gates that it connects to as an input wire. This is possible since $g$ is not in the output layer  of the circuit. We assign type set ${\Rset}$ to gate $g$ and we add $g$ to set $R$. Notice that $g$ is chosen in such a way that the invariant again holds. 

We need to check the effect of pushing possible negations through the output wire of $g$. Each of the gates that the output wire connects to (it may have fanout $>1$) is not in  $R$ (by our invariant) and such gates have still type set ${\Lset}$ assigned to them.
Such a gate can therefore absorb negations that come over its input wires  because 
$\leftindex_{\circ}{\underline{{\Lset}}}_\circ  = {\Lset}$. This shows that after re-assigning $g$ to type set ${\Rset}$, we still obtain a functionally equivalent circuit.


Assume that there only exist gates $g\not\in R$  outside the primary output layer that have at least one of its input wires not from a gate in $R$ and not from an actual input to the circuit. Then that input wire connects as an output wire to another gate $g'\not\in R$ which is outside the primary output layer. According to our assumption, this gate $g'$  connects through one of its input wires to another gate $g''\not\in R$ outside the primary output layer, and so on. Since the circuit has no loops and is finite, this leads to a contradiction. We conclude that our assumption is wrong. Therefore, as long as there exists a gate $g\not\in R$ outside the primary output layer, then there also exists a gate $g\not\in R$ outside the primary output layer with each of its input wires either from $R$ or from an actual input to the circuit. 
This means that when our R-Wave simplification ends,  there are no more  gates $g\not\in R$ outside the primary output layer and
%
we have arrived at an equivalent circuit where each gate that is not part of the primary output layer is in ${\Rset}$.


\subsection{Proof of Theorem~\ref{thm:sz}}


\noindent{\bf S-Class Simplification:}
An S-Class gate is a non-input-layer gate whose two input wires each originate from a source with fanout = 1.  To analyze their interconnections, we form a directed graph whose nodes are exactly the S-Class gates.  We draw an edge from gate $g$ to gate $h$ if and only if the output wire of a fanout = 1 gate $g$ feeds into $h$. Because the original circuit's graph is acyclic, this induced subgraph on S-Class gates is also acyclic. Hence, this graph naturally decomposes into a forest of directed trees, each tree forming an S-Class subtree of connected S-Class gates (which may be a lone node/gate, a linear chain, or a branching binary tree) within the larger circuit structure. In this S-Class graph, each node/gate has at most one outgoing edge and up to two incoming edges from other S-Class gates. Nodes without any outgoing edge serve as the roots of their belonging subtree. In this section, we analyze how the type sets for gates in an S-Class subtree can be further simplified.

Let us consider one of the S-Class subtree. We first consider its root. The root gate starts with type ${\Lset}$. 
\begin{align}
{\Lset} &= \leftindex_{\circ}{\underline{\{\gate{XOR}\}}}_\circ  
           \cup \leftindex_{\circ}{\underline{\{\gate{AND}\}}}_\circ 
           \cup \leftindex_{\circ}{\underline{\{\gate{NAND}\}}}_\circ \nonumber \\
        &\quad \cup \{ \gate{TRUE}, \mathsf{NOT \ A}, \gate{NOT \ B}\}
           \cup \{ \mathsf{FALSE}, \mathsf{A}, \gate{B}\} \nonumber \\[4pt]
&= \leftindex_{\circ}{\underline{\{\gate{XOR}\}}}_\circ  
   \cup \leftindex_{\circ}{\underline{\{\gate{AND}\}}}_\circ 
   \cup \leftindex_{\circ}{\underline{\{\gate{NAND}\}}}_\circ  \nonumber \\
&\quad \cup  \leftindex_{T}{\underline{\{\gate{AND}\}}}_T 
   \cup \leftindex_{T}{\underline{\{\gate{NAND}\}}}_T \nonumber \\[4pt]
&= \leftindex_{\circ}{\underline{\{\gate{XOR},\gate{AND},\gate{NAND}\}}}_\circ 
   \cup \leftindex_{T}{\underline{\{\gate{AND},\gate{NAND}\}}}_T.
\label{Ldecomposition}
\end{align}

From decomposition (\ref{Ldecomposition}), we infer that we may restrict the root gate to type set
$$ {\Sset} = \{ \gate{XOR}, \gate{NAND}, \gate{AND} \}$$
if we appropriately modify the gates $g_a$ and $g_b$ which provide output wires with fanout=1 that serves as input wires $\mathsf{A}$ and $\mathsf{B}$, respectively, to the root gate.
These modifications for $g_a$ and $g_b$ can be no change to the type, a negation of the type, or replacing the type with $\gate{TRUE}$.

Now we consider the non-root gates of the subtree. Our aim is to restrict their types preferably to type set ${\Sset}$.
Initially, we already accomplished this for the root gate.

Let $Q$ be a subtree of our subtree including the root gate. As an induction hypothesis we have that all the gates in $\mathcal{Q}$ are restricted to ${\Sset}$. We start with $Q$ consisting only of the root gate -- our induction hypothesis holds. We extend $Q$ with a new gate $g$ from our subtree. We remind the reader that $g$ is in the type set ${\Lset}$ initially. After applying the appropriate changes required by its preceding gate, which is in ${\Sset}$, $g$ will still be within the type set ${\Lset}$. Thus, we can reapply the decomposition (\ref{Ldecomposition}) of type set ${\Lset}$, and reassign the type of $g$ to type set ${\Sset}$ if we appropriately modify its input gates. By induction, we have reassigned the types of all the gates in the subtree to ${\Sset}$. Also, the gates that provide input wires with fanout=1 to the subtree remain in ${\Lset}$ of 16 types to these gates after the required changes from the subtree.

\vspace{.2cm}
\noindent{\bf Z-Class simplification:}  Z-Class gates are defined as those gates that are not part of the first input layer of the circuit and have exactly one input wire with fanout=1 (the other input wire has fanout $>1$). We create a directed acyclic graph where the nodes represent all Z-Class gates and where the directed edges from one node/gate to another node/gate are output wires with fanout=1. By the definition of Z-Class gates, each Z-Class gate can have at most 1 incoming wire from another Z-Class gate and can have at most 1 outgoing wire to another Z-Class gate. The circuit topology has no loops, and therefore our directed graph is a union of disconnected line graphs (and is acyclic). Each line graph has its edges pointing from a bottom node to a root node; their directions are all lined up.
Within the circuit topology, the line graph zigzags; sometimes an edge represents an output wire that is equal to a left input wire, and other times an edge represents an output wire that is equal to a right input wire.
Notice that zigzag line graphs and subtrees do not intersect because Z-Class gates are different from S-Class gates.

Because Z-Class gates never overlap with S-Class gates, and because after S-Class simplification every non-subtree gate still retains the full type set $\mathcal{L}$, it follows that the root of any Z-Class zigzag chain must itself belong to $\mathcal{L}$.

\begin{align}
{\Lset} &= \underline{\gate{XOR}}_\circ \cup 
           \underline{\gate{AND}}_\circ \cup    
           \underline{\gate{NAND}}_\circ \cup 
           \underline{\mathsf{NOR}}_\circ \cup \nonumber \\
        &\quad \underline{\gate{OR}}_\circ \cup 
           \{ \gate{TRUE}, \mathsf{NOT \ A}, \gate{NOT \ B} \} 
           \cup \{ \mathsf{FALSE}, \mathsf{A}, \gate{B} \} \nonumber \\[6pt]
&= \underline{\gate{XOR}}_\circ \cup 
   \underline{\gate{AND}}_\circ \cup    
   \underline{\gate{NAND}}_\circ \cup 
   \underline{\mathsf{NOR}}_\circ \cup \nonumber \\
&\quad \underline{\gate{OR}}_\circ \cup 
   \{ \gate{TRUE}, \mathsf{FALSE} \} 
   \cup \underline{\{\gate{B}\}}_\circ 
   \cup \underline{\{\gate{AND}\}}_T 
   \cup \underline{\{\gate{NAND}\}}_T \nonumber \\[6pt]
&= \underline{\{\gate{XOR}, \gate{AND}, \gate{NAND}, 
     \mathsf{NOR}, \gate{OR}, \gate{B}\}}_\circ 
     \cup \{ \gate{TRUE}, \mathsf{FALSE} \} \nonumber \\
&\quad \cup \underline{\{\gate{AND}, \gate{NAND}\}}_T \nonumber \\[6pt]
&= \underline{\{\gate{XOR}, \gate{AND}, \gate{NAND}, 
     \mathsf{NOR}, \gate{OR}, \gate{B}\}}_\circ \nonumber \\
&\quad \cup \underline{\{\gate{OR}, \mathsf{NOR}, 
     \gate{AND}, \gate{NAND}\}}_T.
\label{zigzagright}
\end{align}

and similarly,
\begin{align}
{\Lset} &= 
\leftindex_{\circ}{\underline{\{\gate{XOR}, \gate{AND}, \gate{NAND}, \mathsf{NOR}, \gate{OR}, \mathsf{A}\}}} \nonumber \\
&\quad \cup 
\leftindex_{T}{\underline{\{\gate{OR}, \mathsf{NOR}, \gate{AND}, \gate{NAND}\}}}.
\label{zigzagleft}
\end{align}

From the decompositions (\ref{zigzagright}) and (\ref{zigzagleft}) of ${\Lset}$ we replace the type set of $g_b$ (or $g_a$ if the root gate has a left input wire with fanout=1) with 
\begin{eqnarray*}
   {\Zset}_{\mbox{\tt right}}&=&\{ \gate{XOR},\gate{AND},\gate{NAND}, \mathsf{NOR},\gate{OR}, \gate{B}
   \} \mbox{ or} \\
      {\Zset}_{\mbox{\tt left}}&=&\{ \gate{XOR},\gate{AND},\gate{NAND}, \mathsf{NOR},\gate{OR}, \mathsf{A}
      \}
\end{eqnarray*}
depending on whether $g_b$ has a right input wire with fanout=1 or a left input wire with fanout=1. Both these type sets have size 6. 



We may now continue reasoning in the way we have done for the S-Class subtree simplification and assign all the nodes in the Z-class zigzag line to the type set ${\Zset}_{\mbox{\tt right}}$ or ${\Zset}_{\mbox{\tt left}}$, and the gates that provide inputs to the last node of the Z-class zigzag line will stay in $\Lset$ after having adapted one of the required changes: no change of its type, a negation of its type, or a $\gate{TRUE}$ assignment.    

When assigning these type sets to the last node, then the reasoning of the S-Class subtree simplification assumes that it connects through its input wire with fanout=1 to a gate with type set ${\Lset}$, which contains the type $\gate{TRUE}$ so that it can absorb the needed $\gate{TRUE}$ assignments to that input wire. However, the above Z-Class simplification can connect through its input wire with fanout=1 of the last node to either a gate with type set ${\Lset}$ or to a subtree root which has type set ${\Sset}$ and does not contain $\gate{TRUE}$. This is not a cause for concern because the subtree as a whole, together with the gates that provide input over the fanout=1 input wires to the subtree does allow the $\gate{TRUE}$ gate functionality by setting all those gates that provide input to the subtree to type $\gate{TRUE}$. This is possible if those gates all have type set ${\Lset}$. However, such a gate may be a root of a zigzag line graph. In the latter case we cannot directly assign this to the $\gate{TRUE}$ type. However, we may follow this zigzag line graph again to its last node. And again, we may enter a subtree and so on. At some moment, we reach the primary input layer of the circuit with its gates (by definition not S-Class or Z-Class gates) all assigned to type set ${\Lset}$ which contains $\gate{TRUE}$. 

\end{document}